\documentstyle[psfig,draft]{mn}

\def\gs{\mathrel{\raise0.35ex\hbox{$\scriptstyle >$}\kern-0.6em
\lower0.40ex\hbox{{$\scriptstyle \sim$}}}}
\def\ls{\mathrel{\raise0.35ex\hbox{$\scriptstyle <$}\kern-0.6em
\lower0.40ex\hbox{{$\scriptstyle \sim$}}}}

\title[A UKIRT/HST Survey for Gravitationally Lensed EROs]
{A \textit{Hubble Space Telescope} Lensing Survey of 
X-ray Luminous Galaxy Clusters:  II. \newline
A Search for Gravitationally Lensed EROs}

\author[Smith et al.]
{Graham P.\ Smith,$^{\! 1}$\footnotemark[1] Ian Smail,$^{\! 1}$
J.-P.\ Kneib,$^{\! 2}$ O.\ Czoske,$^{\! 2}$ H.\ Ebeling,$^{\! 3}$
\and A.C.\ Edge,$^{\! 1}$ R.\ Pell\'o,$^{2}$ R.\,J.\
Ivison,$^{\! 4}$ C.\ Packham,$^{\! 5}$ J.-F.\ Le~Borgne$^{2}$
\vspace*{1mm}\\
$^1$ Department of Physics, University of Durham, South Road,
Durham DH1 3LE\\
$^2$ Observatoire Midi-Pyr\'en\'ees, 14 Avenue E.\ Belin, 31400
Toulouse,
France\\
$^3$ Institute for Astronomy, University of Hawaii, 2680
Woodlawn
Drive, Honolulu HI\,96822, USA\\
$^4$ Department of Physics \& Astronomy, University College
London, Gower Street, London WC1E 6BT\\
$^5$ Department of Astronomy, University of Florida,
Gainesville, FL 32611 USA\\
}

\date{\fbox{\sc Accepted: 24 September 2001}}
\pagerange{000--000}

\begin{document}

\maketitle

\begin{abstract}

We present the results of a survey for Extremely Red Objects (EROs)
undertaken in the fields of ten massive galaxy cluster lenses at
$z\sim 0.2$, combining sensitive, high-resolution {\it Hubble Space
Telescope} imaging with deep, half-arcsecond $K$--band imaging from
UKIRT. We detect 60 EROs with $(R-K)\geq 5.3$, of which 26 have
$(R-K)\ge6.0$ in a total image plane survey area of 49\,arcmin$^2$
down to $K=20.6$, including one multiply-imaged ERO.  We use detailed
models of the cluster lenses to quantify the lens amplification and
thus correct the observed number counts and survey area for the
effects of gravitational lensing.  After making these corrections, we
estimate surface densities at $K\le21.6$ of ($2.5\pm0.4$) and
($1.2\pm0.3$) arcmin$^{-2}$ for EROs with $(R-K)\geq 5.3$ and 6.0
respectively.  These ERO number counts agree with previous shallower
surveys at $K\ls19$ and flatten significantly at magnitudes fainter
than $K\sim19$--20.  This flattening may be due to a transition from
an ERO population dominated by evolved galaxies at $z\sim1$--2
($K\ls19.5$) to one dominated by dusty starburst galaxies at $z>1$
($K\gs19.5$). We also compare our results with various model
predictions, including a model that attempts to explain EROs in terms
of a single population of elliptical galaxies formed at high redshift.
We find that a formation epoch of $z_f\sim2.5$ for this population
matches the observed surface density of $(R-K)\ge5.3$ EROs quite well,
and the $(R-K)\ge6.0$ sample less well.  More sophisticated models,
including semi-analytic prescriptions, under-predict the ERO surface
density by approximately an order of magnitude, suggesting that these
models produce insufficient stars and/or dust at high redshift.  This
deficit of EROs appears to be a general problem with models that
reproduce the median redshift from $K$--selected redshift surveys.
One possible explanation is that the current $K$-- selected redshift
distribution may be incomplete beyond $z\sim1$.

\end{abstract}


\section{Introduction}

\footnotetext[1]{E-mail: graham.smith@durham.ac.uk}

Extremely Red Objects (EROs) are defined by their very red
optical/near-infrared colours, e.g. $(R-K)\ge5.3$ (e.g.\ Moriondo et
al.\ 2000) and typically have $K$--band magnitudes of $K\gs17.5$. Since their
discovery (Elston et al.\ 1988, 1989, 1991; McCarthy et al.\ 1992;
Eisenhardt \& Dickinson 1992; Graham et al.\ 1994; Hu \& Ridgway 1994;
Dey et al.\ 1995), they have intrigued cosmologists, 
but until recently their extreme properties have escaped detailed, 
systematic investigation.

The variety of colour criteria used to define EROs e.g.  $(R-K)\geq5$,
$\geq5.3$, $\geq6.0$ or $(I-K)\geq 4$, $\geq5$, together with
different survey depths and apparently strong clustering (Daddi et
al.\ 2000a) have led to a wide dispersion in the claimed surface
densities of EROs (e.g.\ Hu \& Ridgway 1994; Cowie et al.\ 1994;
Arag\'on-Salamanca et al.\ 1994; Dey et al.\ 1995; Moustakas et al.\
1997; Yamada et al.\ 1997; Beckwith et al.\ 1998; Eisenhardt et al.\
1998; Barger et al.\ 1999; Thompson et al.\ 1999; Daddi et al.\ 2000a;
McCracken et al.\ 2000). The larger and newer surveys suggest that to
$K\ls19$ the surface density of $(R-K)\geq6.0$ galaxies is
$0.17\pm0.07$ arcmin$^{-2}$ (Daddi et al.\ 2000a), increasing to
$0.28\pm0.04$ arcmin $^{-2}$ at the faintest limit probed to date 
$K\ls20$ (Thompson et al.\ 1999).

The optical faintness of EROs ($R\gs24$) and the lack of near-infrared
multi-object spectrographs have combined to severely limit the number
of spectroscopic identifications. However, the available
spectra indicate that EROs comprise two broad classes of system:
galaxies with evolved stellar populations at $z\gs1$ (e.g.\ Dunlop
et al.\ 1996; Soifer et al.\ 1999; Liu et al.\ 2000) and dust-reddened
starbursts at similar and higher redshifts (Graham et al.\ 1996; Dey
et al.\ 1999). A few starburst EROs have also been detected in radio and 
sub-millimetre (sub-mm) wavelength surveys of the distant
universe (Spinrad et al.\ 1997; Smail et al.\ 1999a, 2000; Dey et al.\
1999; Gear et al.\ 2000; Barger et al.\ 2001). However, morphological
studies (e.g. Treu et al.\ 1999; Moriondo et al.\ 2000) suggest that
at $K\sim19$ the bulk of EROs are evolved elliptical galaxies.

These two classes of EROs may represent different phases in the
formation and evolution of a single family of high redshift galaxies:
massive ellipticals. However, rival galaxy formation theories
(pure luminosity evolution and hierarchical clustering) predict very 
different formation epochs for such galaxies. The former predicts
formation at high redshift in what is traditionally known as the
``monolithic collapse'' scenario (e.g.\ Eggen et al.\ 1962; Larson et
al.\ 1975; Tinsley \& Gunn 1976) followed by subsequent passive
evolution. The latter predicts formation through the merging of disk
galaxies at $z\ls1$ (e.g.\ White \& Frenk 1991; Kauffmann et al.\
1993, 1996; Baugh et al.\ 1996; Cole et al.\ 1996, 2000). The
properties of EROs, specifically, their redshift distribution and
spectral characteristics, are therefore a crucial test of galaxy
formation theories and will improve our ability to make complete
measurements of the star formation history of the universe (e.g.\
Madau et al.\ 1996; Smail et al.\ 2001).

Many groups are currently investigating different aspects of the ERO 
population using a variety of
observational approaches, for example: wide-field imaging surveys
(Daddi et al.\ 2000a; Thompson et al.\ 1999), morphological studies (Treu
et al.\ 1999; Moriondo et al.\ 2000), searches for over-densities 
around radio-loud
AGN (Willott et al.\ 2000; Chapman et al.\ 2000; Cimatti et al.\
2000), deep $K$--band number counts (McCracken et al.\ 2000; Barger et
al.\ 1999), searches for sub-mm counterparts (Smail et al.\ 1999a),
photometric redshift studies (Fontana et al.\ 1999), studies of faint
field spheroidals (Menanteau et al.\ 1999) and high redshift hosts of
optically faint X-ray sources (Cowie et al.\ 2001).

This paper describes a survey whose aim is to produce a sample of EROs
which is optimised for follow-up, with the specific goal of improving
the opportunities for measuring redshifts and spectral
properties. To achieve this we take advantage of the natural
magnification of massive clusters of galaxies to amplify background
source populations, in this case EROs, and so boost their apparent
brightness. This approach has been successfully applied to the
investigation of other faint sources with steep number counts, such as
mid-infrared and sub-mm galaxies (e.g.\ Altieri et al.\ 1999; Smail,
Ivison \& Blain 1997) and may similarly aid progress in understanding
the properties of EROs.

We use a unique sample of X-ray luminous galaxy clusters at moderate
redshifts which have been uniformly imaged with the {\it Hubble Space
Telescope (HST)} through the F702W filter to provide deep, high
resolution imaging of the cluster cores -- including many
gravitationally lensed arcs. We have supplemented these deep optical
images with high quality ground-based near-infrared imaging with UKIRT
and we use this combined dataset to construct a sample of
gravitationally amplified EROs.

In the next section we describe the optical and near-infrared imaging
and its reduction. \S3 describes the analysis of these images and the
properties of the ERO sample constructed from them. Finally in \S4 we
discuss our results and list the main conclusions of this
work. Throughout this paper we assume $H_0=50$kms$^{-1}$Mpc$^{-1}$,
$\Omega_0$=0.3 and $\Lambda_0=0.7$.

\section{Observations}

\subsection{\textit{HST\,/\,WFPC2} Optical Imaging}

The {\it HST} imaging of the clusters used in this paper was obtained
as part of a survey of gravitational lensing by X-ray luminous
clusters (Smith et al.\ 2001a, 2001c). The complete sample consists of
twelve X-ray luminous clusters ($L_{\rm X}\ge
8\times10^{44}$\,erg\,s$^{-1}$, 0.1--2.4\,keV) in a narrow redshift
slice at $z\sim 0.2$, selected from the XBACs sample (Ebeling et al.\
1996). For the purposes of this paper we concentrate on the ten
clusters listed in Table~1 for which homogeneous F702W imaging is 
available (the remaining two clusters which are excluded from
our analysis are A\,1689 and A\,2390 for which only archival $F814W$
observations are available).

%
%
\setcounter{table}{0}
\begin{table*}
\caption{ Details of F702W and $K$--band Observations Used to Construct 
the ERO Sample}
\begin{tabular}{lccccccccccrr}
\noalign{\smallskip}\hline
\noalign{\medskip}
Cluster & Redshift & $E(B-V)$ & F702W & \multispan7{\hfil 
$K$--band \hfil} & \multispan2{\hfil N$_{{\rm ERO}}$ \hfil } \cr
& & &  & \multispan2{\hfil WF2 \hfil} & \multispan2{\hfil WF3 \hfil} &
\multispan2{\hfil WF4 \hfil } & Depth  & \multispan2{\hfil 
$(R-K)\ge$ \hfil } \cr
& & & T$_{F702W}$ & T$_K$ & FWHM & T$_K$ & FWHM & T$_K$ & FWHM & $K$($80\%$) & 
$5.3$ & $6.0$ \cr
& & & (ks) & (ks) & ($''$) & (ks) & ($''$) & (ks) & ($''$) & & & \cr
\noalign{\smallskip}\hline
\noalign{\smallskip}
\multispan3{UKIRT Observations \hfil }\cr
\noalign{\smallskip}
A\,68 & 0.255 & 0.093 & ~7.5 & 8.8 & 0.43 & 8.0 &
0.48 & 7.5 & 0.50 & 20.51 & 9 & 5 \cr
A\,209 & 0.209 & 0.019 & ~7.8 & 6.5 & 0.53 & 6.5 &
0.51 & 6.5 & 0.38 & 20.42 & 7 & 2 \cr
A\,267 & 0.230 & 0.025 & ~7.5 & 6.5 & 0.37 & 6.5 &
0.40 & 4.4 & 0.38 & 20.69 & 12 & 5 \cr
A\,383 & 0.187 & 0.033 & ~7.5 & 6.5 & 0.59 & 6.7 &
0.57 & 6.5 & 0.42 & 20.65 & 11 & 6 \cr
A\,773 & 0.217 & 0.015 & ~7.2 & 6.5 & 0.65 & 6.5 &
0.55 & 9.7 & 0.67 & 20.52 & 1 & 0 \cr
A\,963 & 0.206 & 0.015 & ~7.8 & 6.5 & 0.59 & 6.5 &
0.52 & 6.5 & 0.70 & 20.64 & 12 & 6 \cr
A\,1763 & 0.228 & 0.009 & ~7.8 & 6.5 & 0.58 & 6.5 &
0.55 & 6.5 & 0.69 & 20.65 & 5 & 2 \cr
A\,1835 & 0.253 & 0.030 & ~7.5 & 6.5 & 0.48 & 6.5 &
0.44 & 6.5 & 0.45 & 20.58 & 1 & 0 \cr
A\,2219 & 0.228 & 0.024 & 14.4 & 6.5 & 0.60 & 9.7 &
0.60 & 6.5 & 0.51 & 20.67 & 2 & 0 \cr
\noalign{\smallskip}
\multispan3{WHT Observations \hfil }\cr
\noalign{\smallskip}
A\,2218 & 0.171 & 0.024 & ~6.5 & \multispan6{\hfil
8.3\,ks~~~0.75$''$ \hfil } & 21.50 & 0 & 0 \cr
\noalign{\smallskip}\hline
\noalign{\smallskip}
\end{tabular}
\end{table*}

Of these ten clusters, A\,2218 was observed in Cycle~4, A\,2219 in
Cycle~6 and the remaining eight clusters in Cycle~8. All clusters
except A\,2219 were observed for three, single orbit exposures with
the {\it WFPC2} camera through the F702W filter; A\,2219 was observed
for six orbits using the same instrument/filter combination.  The
total exposure times for the fields, T$_R$, are listed in
Table~1. Each exposure was shifted relative to the others by ten WFC
pixels ($\sim1.0''$) providing a partial overlap of the WFC
chips. After pipeline processing, standard {\sc iraf/stsdas} routines
were used to align and combine the frames to remove both cosmic rays
and hot pixels. Corrections for under-sampling of the point spread
function (PSF) and geometric distortion of the optics were made using
the {\sc dither} package within {\sc iraf} (Fruchter \& Hook
1997). More details are given in Smith et al.\ (2001a, 2001c).

We adopt the photometric system from Holtzman et al.\ (1995) and, in
the following, have chosen to convert the $R_{702}$ photometry to
Cousins $R$--band to aid in comparing our results with previous
surveys.  To achieve this we adopt a $(V-R)$ colour for an Sbc galaxy
at $z=1$--2 of $(V-R)\sim 1.1$, based on spectral templates derived
from local galaxies.  This translates into a correction of $R-R_{\rm
702}\sim 0.4$ with a systematic uncertainty of $\ls0.1$ magnitudes
arising from the likely presence of both more and less evolved
galaxies than the adopted Sbc spectral type.  The equivalent
correction for galaxies with the $(V-R)$ colours of cluster
ellipticals at $z\sim0.2$, $(V-R)\sim0.9$, is $R-R_{\rm 702}\sim 0.3$,
with an estimated systematic uncertainty of $\ls 0.05$.  Both of these
corrections are sufficiently accurate to define a sample of EROs in
these fields.

The final WFPC2 frames have an effective resolution of 0.15$''$ and a
typical 3--$\sigma$ detection limit within our 2$''$--diameter
photometry aperture of $R\sim26.6$.

\subsection{Near-infrared Imaging}

The near-infrared ($K$--band) imaging essential for identifying EROs
was obtained for nine of the ten clusters with the UFTI imager on the
3.8--m United Kingdom Infrared Telescope (UKIRT)\footnote{UKIRT is 
operated by the
Joint Astronomy Centre on behalf of the Particle Physics and Astronomy
Research Council of the United Kingdom.}. UFTI incorporates a 1024$^2$
Hawaii-1 detector providing a 92$''$ field of view with 0.0908$''$
pixel$^{-1}$ sampling -- necessary to sample the best seeing provided
by UKIRT, $\ls 0.2$--0.3$''$. The $K$--band imaging of these fields
was obtained in two observing runs in 2000 April 4--7 and September
26--29. Conditions during both observing runs were good, with
reasonable seeing and transparency. Each {\it WFPC2} field was covered
in three pointings (one per WFC chip: WF2, WF3 and WF4), although the
different roll-angles for the various {\it WFPC2} observations means
that a particular near-infrared exposure may overlap more than one WFC
chip. The individual pointings usually consist of $54\times 120$--s
exposures grouped into six sets of nine exposures, or occasionally
$72\times 90$--s exposures in eight sets of nine, each exposure being
offset on a $3\times 3$ grid with 10$''$ spacing, and the origin of
each set was randomly moved by $\sim 3''$. We list the total exposure
times for the individual pointings (T$_K$) and the seeing for each in
Table~1. The median seeing for these 2--hr integrations is 0.5$''$.

The data were reduced using the dedicated UFTI summit data pipeline
(Currie et al.\ 1999) including dark subtraction, flat fielding using
a local median sky, resampling onto an astrometric grid and combining
to remove defects and cosmic ray events. A small amount of the data
was re-reduced in Durham following a similar scheme and using standard
{\sc iraf} tasks. Calibration was obtained from UKIRT Faint Standards
(Casali \& Hawarden 1992) interspersed throughout the science
observations. We estimate that our absolute calibration is good to $\ls
0.05$\,mag. A reddening correction of $E(R-K)=2.31 E(B-V)$, was also
applied to the $(R-K)$ colours following Schlegel et al.\ (1998). The
final UFTI $K$--band images typically reach a median 80\% completeness 
of $K\sim 20.6$ (Table~1 \& \S2.4). For comparison a passively
evolving $L^\ast$ elliptical will have $K\sim18$--19 at $z=1$.  We
show an example of one of our fields in Fig.~1.

The tenth cluster, A\,2218, was observed in commissioning time with
the new INGRID near-infrared imager (Packham et al.\ 2001) on the
4.2--m William Herschel Telescope (WHT)\footnote{Based on observations
made with the William Herschel Telescope operated on the island of La
Palma by the Isaac Newton Group in the Spanish Observatorio del Roque
de los Muchachos of the Instituto de Astrofisica de Canarias}. INGRID
comprises a 1024$^2$ Hawaii-1 detector giving a 248$''$ field of view
with 0.242$''$ pixel$^{-1}$ sampling. These data comprise a 8.3\,ks
exposure of the whole {\it WFPC2} field in A\,2218 (Kneib et al.\
1996) in the $K_s$ band. The seeing on the final stacked frame is
0.75$''$ and the frame reaches an 80\% completeness limit of $K=21.5$ for
point sources. The reduction, calibration and analysis of these data
are discussed in more detail in Smail et al.\ (2001). For uniformity
in our analysis we restrict ourselves to a limit of $K\le20.6$ in
A\,2218.

In addition to the $K$--band imaging described above, a sub-set of the
cluster fields were also observed in the $J$--band.  We observed A\,773, 
A\,963, A\,1763 and A\,1835 with UKIRT/UFTI on the nights of 2001 
February 16 and April 4 in non-photometric conditions and $\sim0.6''$ 
seeing, and A\,68 A\,267 and A\,2219 with WHT/INGRID on the nights of 
2001 June 27 and 2001 May 6, in $\sim0.8''$ seeing and photometric 
conditions (Table~2).  All of these observations
were reduced in a similar manner to the $K$--band images.  The UFTI 
observations reach a 5--$\sigma$ limit of $J\sim21$, whilst INGRID frames 
reach 5--$\sigma$ depths of $J\sim$20--21. Finally, a $J$--band image 
of A\,2218 is also available from the INGRID commissioning observations of 
that cluster, however, as we find no ERO sources brighter than $K=20.6$
within that field, we do not discuss these data further.

\subsection{Ground-based Optical Imaging}

To study the spectral energy distributions of EROs between the $R$--
and $K$-- bands, we also make use of archival $I$--band imaging of the
clusters in our sample. These observations come from a number of
ground-based 4--m class telescopes and comprise 0.5--3.6\,ks
integration in 0.7--1.4$''$ seeing (Table~2), reaching $I\gs23$. We
use panoramic imaging of A\,68, A\,209, A\,267 and A\,383 from the
CFH12k camera on the 3.6--m Canada-France-Hawaii
Telescope,\footnote{Based on observations made with the Canada France
Hawaii Telescope, operated on the island of Hawaii by the Canada
France Hawaii Telescope Corporation} taken on 1999 November
10--14. More details of their reduction and analysis can be found in
Smith et al.\ (2001a, 2001c) and Czoske et al.\ (2001). For A\,773 and
A\,963 we use $I$--band images taken on 1994 December 8--9 with PFCCD
of the 4.2--m William Herschel Telescope\footnote{Based on
observations made with the William Herschel Telescope operated on the
island of La Palma by the Isaac Newton Group in the Spanish
Observatorio del Roque de los Muchachos of the Instituto de
Astrofisica de Canarias}. Finally, we exploit imaging of A\,1763,
A\,1835, A\,2218 and A\,2219 taken with the COSMIC imager spectrograph
on the Hale 5--m\footnote {This work is based on observations obtained
at Palomar Observatory, which is owned and operated by the California
Institute of Technology}, more details of these observations and
details of their reduction can be found in Smail et al.\ (1998) and
Ziegler et al.\ (2001).

\section{Analysis and Results}

%
%
\begin{table}
\caption{ Summary of $I$-- and $J$--band Observations }
\begin{tabular}{lccccc}
\noalign{\smallskip}\hline
\noalign{\medskip}
Cluster & \multispan2{\hfil $I$--band \hfil} & \multispan3{\hfil 
$J$--band \hfil} \cr
& T$_I$ & FWHM & T$_J$ & FWHM & Field(s) \cr
& (ks) & ($''$) & (ks) & ($''$) & Observed$^a$ \cr
\noalign{\smallskip}\hline
\noalign{\smallskip}
A\,68 & 3.6 & 0.7 & 3.2 & 0.8 & WFPC2 \cr
A\,209 & 3.6 & 0.7 & ... & ... & ... \cr
A\,267 & 0.9 & 0.7 & 1.3 & 0.8 & WFPC2 \cr
A\,383 & 3.6 & 0.7 & ... & ... & ... \cr
A\,773 & 0.9 & 1.4 & 3.2 & 0.6 & WF4 \cr
A\,963 & 0.6 & 1.1 & 3.2 & 0.6 & WF2/3/4 \cr
A\,1763 & 0.5 & 1.1 & 3.2 & 0.6 & WF2/4 \cr
A\,1835 & 1.0 & 1.1 & 3.2 & 0.6 & WF2 \cr
A\,2218 & 21.7 & 0.9 & 6.5 & 0.8 & WFPC2 \cr
A\,2219 & 0.5 & 1.1 & 2.2 & 0.8 & WFPC2 \cr
\noalign{\smallskip}\hline
\noalign{\smallskip}
\end{tabular}
\newline
{\small (a) The $I$--band observations cover the whole WFPC2 
field of view.  For the $J$--band: WFPC2 denotes WHT/INGRID 
observations which cover the whole field; WF2/3/4 indicate 
UKIRT/UFTI imaging of a particular WFC chip.}
\end{table}

\subsection{Source Detection and Photometry}

%
%
\begin{figure*}
\psfig{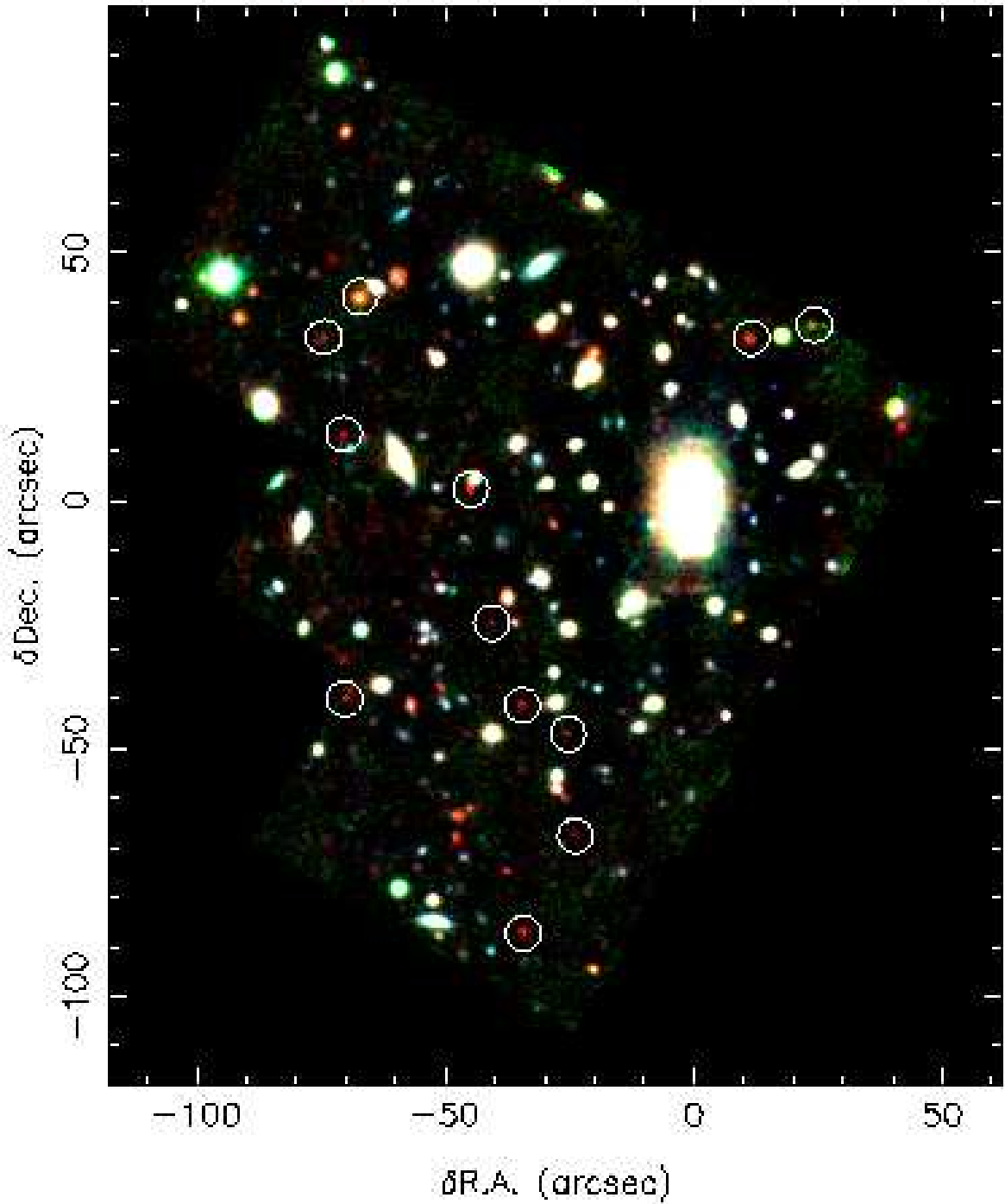}
\vspace{2cm}
\caption{An example cluster field from our {\it HST}\,/\,UKIRT survey,
showing a true colour $RIK$ image of the core of A\,963. The circles
indicate the thirteen EROs with $(R-K)\geq5.3$ detected in this
field. The two EROs north-west of the cD appear to be elliptical
galaxies with similar redshifts on the basis of their
optical/near-infrared colours and morphology (\S3). The field is
centred on the cluster cD and has North top and East left. Note the
very blue giant arc visible south of the cD.}
\end{figure*}

The near-infrared images of each cluster were aligned and mosaiced
together (if necessary) to create one $K$--band frame per cluster that
covers the relevant {\it WFPC2} field of view. The {\it WFPC2} frames
were rotated and aligned to the $K$--band frames with an rms tolerance
of $\ls 0.01''$ and the whole field astrometrically calibrated using
the APM catalogue to an absolute accuracy of $0.4''$.

To produce a catalogue of EROs in these fields we first analysed the
$K$--band frames using the SExtractor package (Bertin \& Arnouts
1996). All objects with isophotal areas in excess of 10 pixels
($0.082$\,arcsec$^{2}$) at the $\mu_{K}=23$\,mag\,arcsec$^{-2}$
isophote ($1.5\sigma$\,pixel$^{-1}$) and lying within the WFPC2 
field of view were selected. Across the ten
clusters, this survey covers a total area of 49 arcmin$^2$, excluding
the PC chips, and these catalogues contain a total of 2,382 sources.
We adopt the {\sc mag}\_{\sc best} magnitude computed by SExtractor
as the total $K$--band magnitude of each source.

We perform extensive ($\sim10^4$ realisations) Monte Carlo simulations
to measure the completeness limits of these catalogues by suitably 
scaling and re-inserting a moderately bright ($K\sim19$), compact ERO 
source into the science frames.  The resulting 80\%
completeness limits (roughly equivalent to 5--$\sigma$ detection limits) 
are presented in 
Table~1. We also re-perform these
simulations using a more diffuse galaxy of similar magnitude, and find
that the typical 5--$\sigma$ limiting depth was $\ls0.3$ magnitudes
brighter than that for the compact source.  This difference is not
large enough to have a significant impact on the results presented in
this paper.  

We measured the $(R-K)$ colour of all these sources using
a 2$''$ diameter aperture on seeing matched $R$-- and $K$--band
frames.  As the image smoothing required to match the seeing reduces
the pixel to pixel variation in the sky backgrounds, the noise
estimates from the seeing-matched frames are compromised. We therefore
exploited the overlap regions between the three UFTI pointings used to
cover each {\it HST} field to compare independent photometry of
sources as a function of magnitude, and so derive conservative
estimates of the photometric uncertainties as a function of
magnitude. We applied the same principle in the $R$--band, taking
advantage of the six 2.4\,ks exposures of A\,2219 (\S2.1), to make
independent measurements on two frames, each derived from an
independent sub-set of three exposures.  This provides a reliable and
conservative estimate of the photometric errors of sources in the
$R$--band. We also estimate the 5--$\sigma$ depth of the $R$--band
imaging to be $R=26.0$. However, for the purposes of our final
photometry we adopt a 3--$\sigma$ $R$--band detection limit of
$R=26.6$ to discriminate between $(R-K)\geq 5.3$ and $(R-K)\geq 6.0$
(\S2.4) sources lying close to both detection limits.

We show the $(R-K)$--$K$ colour-magnitude diagrams for the ten
clusters in Fig.~2. The strong sequences of red galaxies identifiable
in all the panels correspond to the evolved early-type cluster
members. These provide a useful check of the calibration of the
$(R-K)$ colours for sources in these fields. We estimate that the rms
scatter in the colours of the cluster early-type sequences at a fixed
luminosity as a function of redshift is $\sim 0.08$\,mags, confirming
that any field-to-field offset between our photometry of galaxies detected 
in each cluster field is $\ls0.08$ magnitudes.

Finally, we measure the $(I-K)$ and $(J-K)$ (where $J$--band imaging
is available) colours of the ERO sample defined in \S3.2 using a 2$''$
diameter aperture on seeing-matched frames, adopting a 3--$\sigma$
detection limit when no $I$-- or $J$-- band counterpart is detected.

%
%
\begin{figure*}
\centerline{\psfig{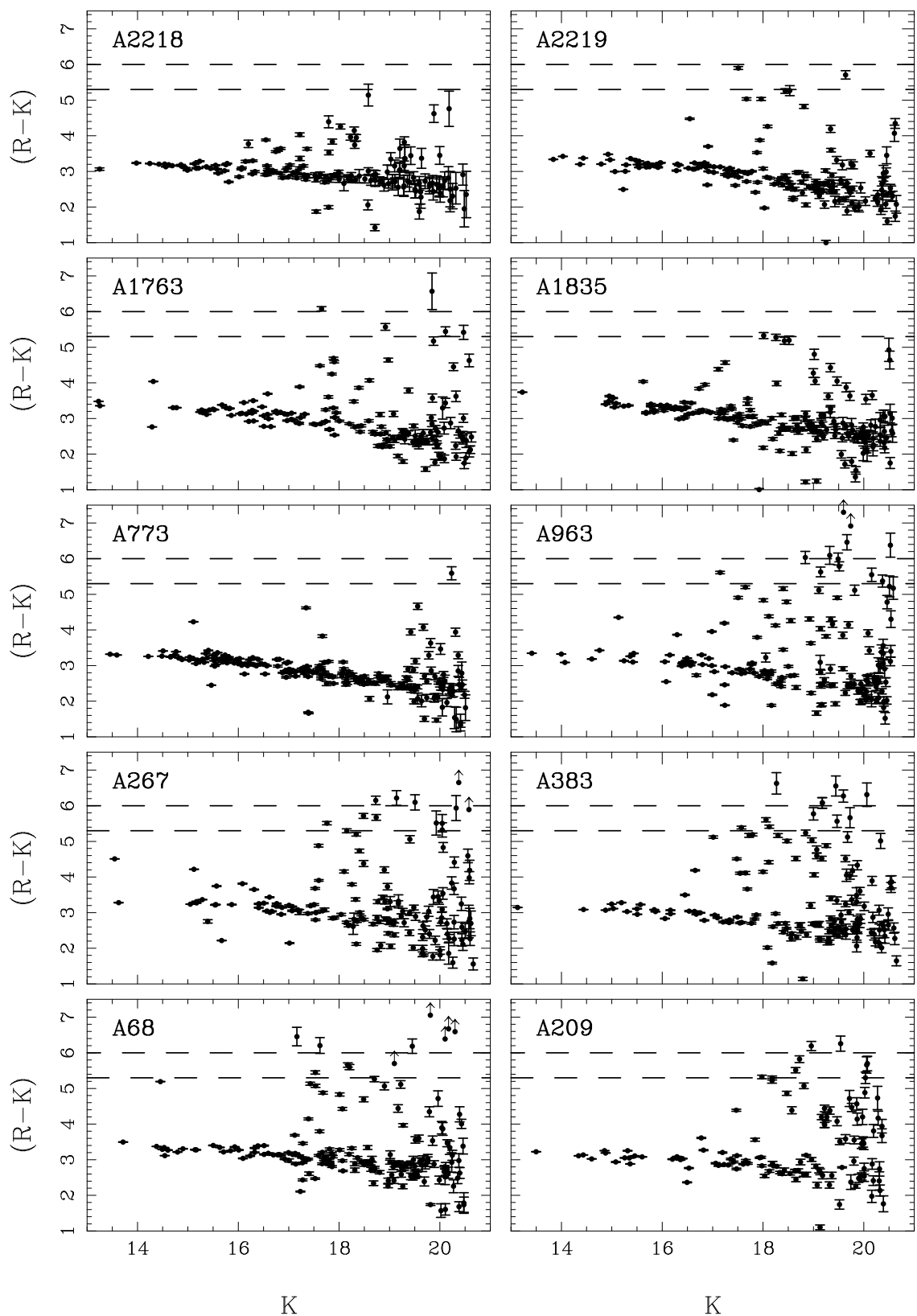}}
\caption{$(R-K)$--$K$ colour-magnitude diagrams for the ten clusters
in our survey, showing the order of magnitude field to field variation
in ERO number counts. The $K$--band magnitudes in this figure are
observed magnitudes (i.e.\ not corrected for gravitational
amplification) and are plotted down to the 80\% completeness 
limits listed in Table~1 (median 80\% completeness limit is $K=20.6$). The
two dashed lines on each panel indicate the $(R-K)\ge5.3$ and 6.0 ERO
selection criteria respectively. Points marked by an upward pointing
arrow are 3--$\sigma$ limits in $(R-K)$.}
\end{figure*}

%
%
\medskip
\begin{figure*}
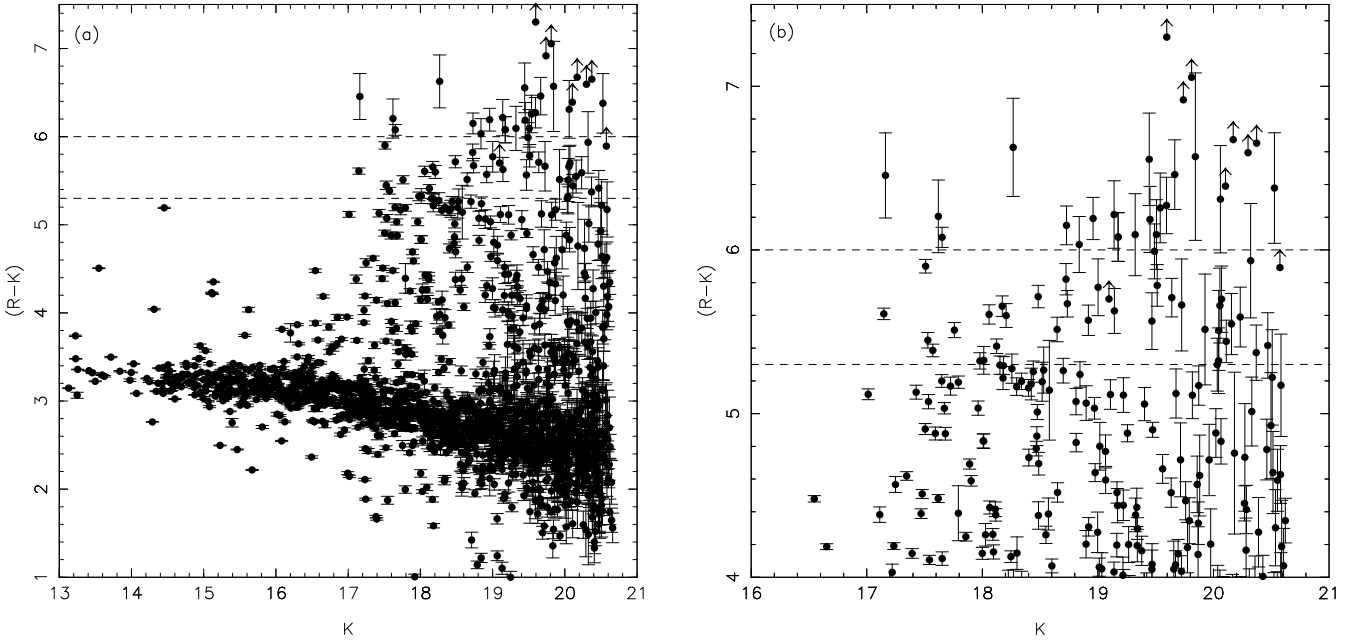

\centerline{\psfig{file=figs/fig3a.ps,angle=-90,width=85mm}
\hspace{5mm}
\psfig{file=figs/fig3b.ps,angle=-90,width=85mm}}
\caption{a) Composite $(R-K)$--$K$ colour-magnitude diagram for the
ten cluster fields in our survey. b) Zoom into $(R-K)>4$ region of the
composite $(R-K)$--$K$ colour-magnitude diagram shown in (a). The two
dashed lines in each panel indicate the $(R-K)\geq 5.3$ and 6.0 ERO
selection criteria. Galaxies marked by an upward pointing arrow are
3--$\sigma$ limits in $(R-K)$.  We have co-added these objects and 
measure the typical
$R$--band magnitude (in a 2$''$ diameter aperture) to be
$R=27.0\pm0.5$. This suggests that the galaxies not detected in the
$R$--band are a continuation of the $(R-K)\ge6.0$ population, rather
than a distinct population of galaxies with much more extreme
colours. Note, the $K$--band magnitudes plotted in this figure are
observed magnitudes (i.e.\ not corrected for gravitational
amplification).}
\end{figure*}
\medskip

\subsection{ERO Selection}

We adopt the definition of an ERO as a galaxy with $(R-K)\geq 5.3$ as
used by Daddi et al.\ (2000a), and also employ a more stringent
definition of $(R-K)\geq6$ (Thompson et al.\ 1999). These sample
boundaries are shown for the individual cluster fields in Fig.~2, we
list the numbers of galaxies in each class in the various clusters in
Table~1 and give a catalogue of the sources in Table~3.

We show in Fig.~3 the composite $(R-K)$--$K$ colour magnitude diagram
for all ten clusters, along with the sample boundaries. In total we
find 60 sources with $(R-K)\geq 5.3$ in the ten fields (an image plane
survey area of 49 arcmin$^2$) and 26 with $(R-K)\geq 6$ down to our
median 80\% completeness limit of $K=20.6$.

In addition to the Monte Carlo simulations described in \S3.1, we
performed two further checks to verify the completeness of our ERO
sample. First, we visually checked true-colour optical/near-infrared
images of the clusters (e.g.\ Fig.~1) to guard against losing ERO
candidates due to contamination of photometric apertures by nearby
sources with less extreme colours. In two cases we re-calculated the
colours of objects after masking the light from nearby galaxies to
ensure a more reliable measurement. Second, we searched for EROs
hidden under the halos of bright cluster ellipticals by subtracting a
median-smoothed version of the each $K$--band frame from the original
frame to produce a ``difference'' frame. This search revealed no 
further EROs.  

These ``difference'' frames contain residual flux from the central 
regions of the bright cluster ellipticals, with the result that a 
fraction of each {\it WFPC2} field of view remains obscured.  We  
estimate this fraction, first thresholding the ``difference'' frames at 
a level where the fainter EROs in our sample would not be detected.  We 
then ray trace these thresholded frames back to the source plane (using 
the lens models described in 
\S3.3).  As the lens amplification is highest at the cluster centre and 
the centre of other bright cluster ellipticals, the transformation to 
the source plane results in the obscured fraction of each field 
of view being typically $\ls5$\%.

We estimate contamination of the sample by red, low mass stars by
visually comparing the $K$--band morphologies of the 19 EROs in our
sample with a FWHM $\le 0.8''$ to the images of
morphologically-classified stars selected from the {\it HST} frames.
Only one of these sources (ERO\,J024805$-$0330.2) has a star-like
morphology; we therefore flag it as a possible star or AGN, but choose
not to remove it from our analysis. This very low level of stellar
contamination ($\le1.5\%$) is consistent with that determined by Daddi
et al.\ (2000a) and Thompson et al.\ (1999).

A further potential bias in colour-selected surveys in gravitational 
lens fields  is differential amplification across a galaxy image, 
causing separate regions of the galaxy with possibly different underlying 
colours to suffer different degrees of amplification.  As the lens 
amplification only varies strongly with position close to critical lines, 
this bias is only a concern for multiply-imaged galaxies.  Only one 
of our sample of 60 EROs is multiply imaged (\S3.7).  We therefore 
conclude that this bias has a negligible effect on our sample.

\subsection{Surface Density of EROs}

In a ``blank'' field survey, the surface density of sources brighter than
a given limiting magnitude can be calculated by dividing the number of
detected sources by the surveyed area. This simple calculation is more
complicated in the field of a gravitational lens, due to the amplification 
of the source flux and the accompanying distortion of the background sky
(source plane). Consequently, some regions of the source plane are
observed to greater depths than others, even if the image plane is
observed to a uniform depth. These effects can be quantified by
constructing a detailed model of the gravitational lens. Such models are
then used to compute the lens amplification as a
function of image plane position. This knowledge of the lens
amplification allows the surface density of sources to be reliably
estimated as described in more detail below. First, we briefly discuss
the lens models used to quantify the lens amplification in our cluster
sample.

Smith et al.\ (2001c; hereafter S01) have constructed a detailed
model of each cluster in our sample using the parametric lens
inversion method of Kneib et al.\ (1996). S01 discuss the robustness
of these lens models in considerable detail. In this paper, we
summarise the redshift information used to constrain the models and
thus provide an overview of the reliability of these models. The most
robust models are those constrained by the spectroscopic redshift of a
multiply-imaged background galaxy. Four of the ten clusters fall into
this category. Half of the remaining six clusters contain faint
multiply-imaged sources, whilst only three contain no obvious multiple
images in our F702W frames.  S01 therefore used both photometric
redshift techniques and uniformly selected samples of weakly sheared
background galaxies to constrain the lens models for these six
clusters.  Consequently, the amplifying power of the four
spectroscopically constrained models is known to $\sim$2--5\%, whilst the
power of the other six is typically known to an accuracy of
$\sim10$\%. We therefore use S01's lens models to remove the effects
of lens amplification from both the source counts and the surveyed 
area using the method described below (Blain et al.\ 1999).

The amplification suffered by each background galaxy depends on its
redshift ($z_{\sc s}$) and the redshift of the intervening lens ($z_{\sc
l}$). This redshift dependence is weak if $z_{\sc s}\gg z_{\sc l}$, and
as the cluster lenses all lie at $z_{\sc l}\sim0.2$ and the background
galaxies are all expected to lie at $z_{\sc s}\sim1$--2, this regime
applies. We therefore adopt a single source plane of $z_{\sc s}$=1.5 and
use each model to compute a map of lens amplification as a function of 
image plane position for this value of $z_{\sc s}$ in each of the ten 
cluster fields. Adoption of a single value of $z_{\sc s}$ introduces 
an uncertainty of $\sim10$--20$\%$ into the final surface density values, 
which is comparable with the Poisson noise in the raw number counts.

We first use the amplification maps to de-amplify the image plane flux of
each ERO in our sample and hence obtain their source plane
$K$--band magnitudes. As gravitational lensing is achromatic,
no correction is required to the $(R-K)$ colour. The number of
EROs that are brighter than a source plane limiting magnitude $K_{\rm
lim}$, $N_{\rm raw}(<K_{\rm lim})$, can then be found by simply counting
the number of sources brighter than $K_{\rm lim}$ in the source plane
after correcting for lensing. A simple Poisson uncertainty is attached to
this value.

We then calculate the area of the background sky within which each ERO is
detectable in the relevant cluster field in the following manner. An ERO
with a source plane magnitude of $K_{\rm source}$ will appear in the image
plane of the relevant cluster with a magnitude brighter than an image
plane detection limit of $K_{\rm det}$ if it is magnified by a factor
greater than $\mu_{\rm min}=10^{-0.4(K_{\rm det}-K_{\rm source})}$. The area
in the source plane within which such a galaxy would be detected in that
cluster is thus $A_{>}(\mu_{\rm min})$, where $A_{>}$ is the area of the
$z_{\rm source}=1.5$ source plane behind each cluster that lies within the
{\it WFPC2} field of view and is magnified by a factor greater than
$\mu_{\rm min}$ (Fig.~4).  Finally, we divide the number of galaxies 
$N_{\rm raw}$ by the sum of the
areas $A_{>}(\mu_{\rm min})$ for all ten clusters, to obtain the
cumulative surface density $N(<K_{\rm lim})\simeq N_{\rm raw}(<K_{\rm
lim})/\Sigma A_{>}(\mu_{\rm min})$.

We present in Fig.~5 the cumulative surface density, $N(\le K)$, of
$(R-K)\ge5.3$ and $(R-K)\ge6.0$ EROs detected in our survey after
correcting the source fluxes and source plane surface areas for lens 
amplification. We also show the results of recent shallower,
wide-field surveys (Daddi et al.\ 2000a; Thompson et al.\ 1999).  The 
three datasets agree in the region of overlap 
($K\sim17.5$--19.5). We estimate the cumulative surface density of
EROs at $K\ls21.6$ with $(R-K)\ge5.3$ to be ($2.5\pm0.4$)\,arcmin$^{-2}$ 
and with $(R-K)\geq6.0$ to be ($1.2\pm0.3$)\,arcmin$^{-2}$.
We note that the $(R-K)\ge6.0$ EROs appear to comprise a constant
fraction of the the overall $(R-K)\ge5.3$ population at all
magnitudes: $\sim0.40\pm0.08$.

The slope of Daddi et al.'s (2000a) cumulative number counts for
EROs with $(R-K)\ge5.3$ is $\alpha=1.05\pm0.05$ at $K<19.5$, where $N(\le
K)=10^{\alpha K}$. However, fainter than $K\sim19.5$, we estimate the
slope of our $(R-K)\ge5.3$ cumulative number counts to be
$\alpha=0.30\pm0.01$, suggesting a break in the surface density of
$(R-K)\ge5.3$ EROs at $K\sim19$--20. Number counts of the $(R-K)\ge6.0$
population reveal a similar break, with slopes of $\alpha=1.43\pm0.25$ 
and $\alpha=0.37\pm0.02$ for brighter and fainter EROs respectively.

%
%
\medskip
\begin{figure}
\centerline{\psfig{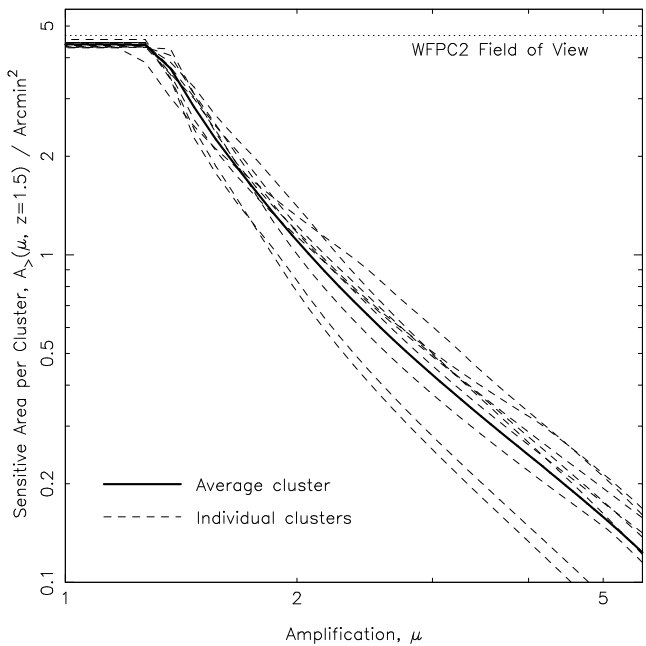}}
\caption{Cumulative area of the source plane in our survey, A$_{>}$,
at $z=1.5$, that experiences magnification greater than $\mu$.  The
similarity of the individual clusters in this figure reflect the
uniform selection criteria for our cluster sample (\S2.1).}
\end{figure}
\medskip

\subsection{Colour Distributions}

We present the $(R-K)$--$(I-K)$ colour-colour diagram of our ERO
sample in Fig.~6a.  We also plot in Fig.~6a the expected colours of 
passively evolving ellipticals as a function of redshift (for a 
formation redshift of $z_f=2.5$ -- see also \S4.1). 
We begin by comparing the $(R-K)$ and $(I-K)$ ERO
selection criteria. Of the $(R-K)\ge5.3$ EROs, $\ls10$\,(15$\%$) have
$(I-K)<4$ and would therefore not be classified as EROs in an $(I-K)$
selected sample.  Most of these galaxies are also blue in $(R-K)$, and 
given their proximity to the $z\ls1$ portion of the evolutionary track,
we suggest that they are probably early-type galaxies whose
4000--${\rm \AA}$ break lies between the $R$-- and $I$--bands.

We also see that a substantial fraction of
the $(R-K)\ge6.0$ galaxies are redder than the passive evolutionary track.
Whilst we treat this comparison with caution, due to uncertainties in
both photometric measurements and calibration onto the models, this
suggests that the $(R-K)\ge6.0$ population may contain a sizeable
fraction of starburst galaxies. We also measure the typical colour of
those EROs for which we can only place a limit on their $(R-K)$ colour
by co-adding the F702W--band images of these EROs. Using the same photometric
approach as in \S3.1, we obtain a typical $R$--band magnitude for
these galaxies of $R=27.0\pm0.5$, just 0.4 magnitudes fainter than the
3--$\sigma$ limit of $R=26.6$ determined in \S3.2. This suggests that
these galaxies are a continuation of the $(R-K)\ge6.0$ population that
is detected in $R$, rather than a distinct population of galaxies with
much more extreme colours. The three galaxies with the most extreme
$(R-K)$ colours are shown in Fig.~7.

%
%
\medskip
\setcounter{figure}{4}
\begin{figure*}
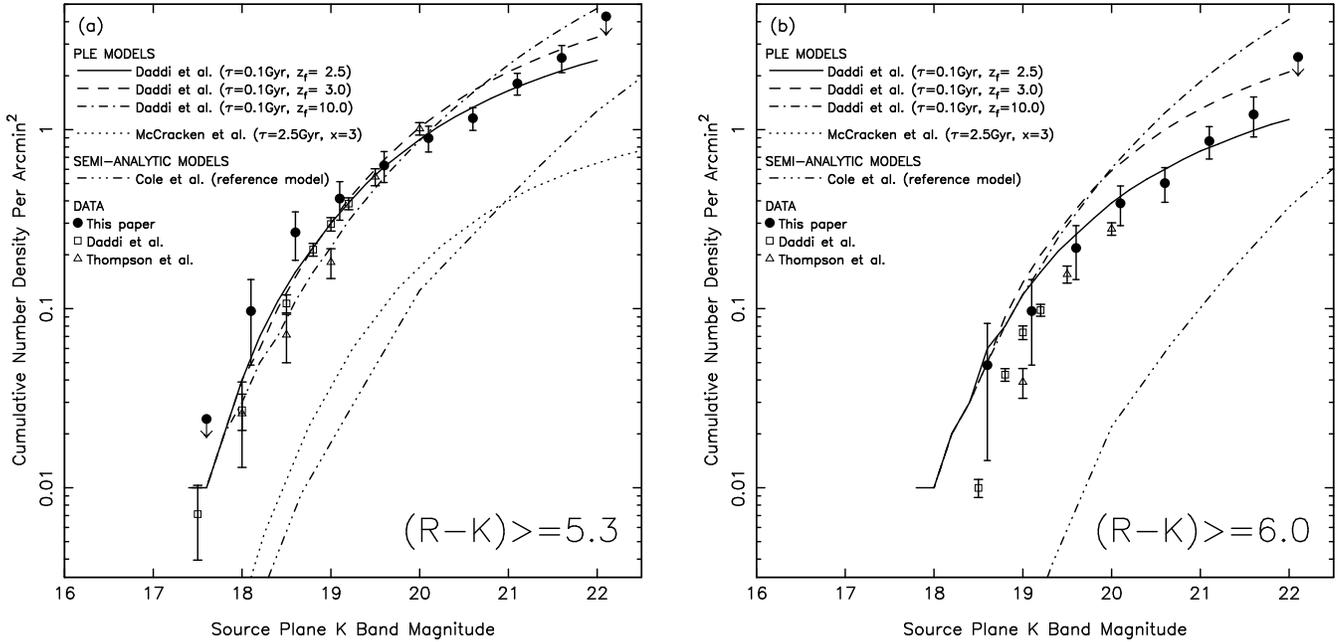

\centerline{\psfig{file=figs/fig5a.ps,angle=-90,width=85mm}
\hspace{5mm}
\psfig{file=figs/fig5b.ps,angle=-90,width=85mm}}
\caption{The cumulative surface density of (a)
$(R-K)\geq5.3$ and (b) $(R-K)\ge6.0$ EROs detected through our cluster
lens sample after removal of gravitational amplification, as a function
of limiting source-plane $K$--band magnitude. 
The faintest two bins in both panels are plotted after allowing for the 
effects of incompleteness.
We make no pass-band corrections between the different $K$--band filters
used in the three surveys as such corrections are $\ls0.1$ magnitudes
(Cowie et al.\ 1994; Persson et al.\ 1998). The error bars are all
Poissonian, and therefore generally underestimate the true error due
to source clustering. 
We also plot the data of Daddi et al.\ (2000a) and
Thompson et al.\ (1999), together with the PLE model predictions of Daddi
et al.\ (2000b) and McCracken et al.\ (2000) and the semi-analytic model
prediction of Cole et al.\ (2000). Note that the McCracken et al.\ model
predicts zero surface density of EROs with $(R-K)\ge6.0$.
The Daddi et al.\ models assume ($\Omega_0=0.3$,
$\Lambda_0=0.7$, $H_0=50$kms$^{-1}$Mpc$^{-1}$), McCracken et al.\ assume
($\Omega_0=0.1$, $\Lambda_0=0$, $H_0=50$kms$^{-1}$Mpc$^{-1}$) and Cole et
al.\ assume ($\Omega_0=0.3$, $\Lambda_0=0.7$, $H_0=70$kms$^{-1}$Mpc$^{-1}$). 
Over the redshift range of interest ($z\sim$1--2), the $\Lambda$CDM cosmology
adopted by Daddi et al.\ and Cole et al.\ is very similar to McCracken et
al.'s open cosmology (Peebles 1984 \& McCracken et al.\ 2000). These
differences in cosmological model therefore do not affect the conclusions
drawn in \S4.1 and \S4.2.}
\end{figure*}
\medskip

Pozzetti \& Mannucci (2000) develop a method for breaking the
``age--dust'' degeneracy between passive and starburst EROs.  They
explore the $(I-K)$--$(J-K)$ and $(R-K)$--$(J-K)$ planes, proposing a
classification scheme based on the steepness of the spectral break
between optical and near-infrared bands at $z=1$--2. We show the
$(R-K)$--$(J-K)$ plane of those EROs for which we have obtained
$J$--band detections in Fig.~6b.  We note that all of the 
morphologically classified ``Compact'' EROs (see \S3.5) lie on the 
``Elliptical'' side of the classification boundary.  The situation 
is less clear for the ``Irregular'' galaxies, where the large 
photometric uncertainties mean that we cannot reliably state that 
the majority of these galaxies lie on the ``Starburst'' side of 
the classification boundary.
More accurate $J$--band photometry and spectroscopic identifications of
the brighter members  are required before 
firm conclusions may be drawn from this analysis.

Whilst the foreground cluster lenses increase the chances of
successful spectroscopic identifications, the same amplifying power
allows us to observe yet fainter EROs ($K\sim20$--21 in the image
plane), for which spectroscopic observations will remain unfeasible.
As a test of the capabilities of photometric redshift measurements for
EROs we use {\sc hyper-z} (Bolzonella et al.\ 2000) to estimate
photometric redshifts for those EROs for which $RIJK$ photometry is
available. We use template spectra corresponding to 51 different ages
for a $\tau=0.1$ Gyr $\mu$--model from Bruzual \& Charlot (1993) to
obtain a likelihood map in dust extinction--redshift (A$_V$--$z$)
space for each galaxy. In most cases, valid solutions are possible for
a wide range of redshifts, however the most likely solution is usually
found around $0.8\le z\le1.5$--2.

Future spectroscopic observations (e.g. Smith et al.\ 2001b) in 
conjunction with $J$-- and
$H$--band photometry of the whole sample will probe the redshift
distribution and star formation histories of our ERO sample in
significantly more detail.

\subsection{Morphological Diversity}

To gain further insight into the diversity of the ERO population we
now attempt to quantify the morphologies of our sample. The optical
imaging from {\it HST} has exquisite spatial resolution ($\sim$0.1$''$),
however EROs are optically faint ($R\gs24$), and therefore suffer from
low signal to noise in optical pass-bands. In contrast, EROs are typically
well detected in the near-infrared and our $K$--band imaging from
UKIRT enjoys superb seeing ($\sim$0.5$''$). We therefore exploit both
datasets in the following analysis.

We first attempt to measure the scale-size of each member in our ERO
sample. We fit a Moffatt profile to the $K$--band image of each ERO
from which we estimate the FWHM, and then crudely correct for the effects
of seeing by subtracting in quadrature the FWHM of the PSF on the frame 
(Table~3). These measurements reveal no trend in intrinsic FWHM versus 
colour or $K$--band magnitude, most likely due to uncertainties in the 
individual measurements.

We therefore attempt to measure the properties (scale-size and central
concentration) of a typical ERO in amplification corrected magnitude
bins: $K\le19$, $19<K\le20$ and $K>20$, co-adding the $K$--band images
of the EROs to obtain one representative galaxy per bin. Using
$\eta\,(\theta)=I(\theta)/\langle I\rangle_{\theta}$ where $I(\theta)$
is the surface brightness at radius $\theta$ and $\langle
I\rangle_{\theta}$ is the mean surface brightness within $\theta$
(Petrosian 1976; Kron 1980), we measure the angular size
$\theta_{0.5}$ of the three co-added galaxies where
$\eta\,(\theta_{0.5})=0.5$ (Bershady et al.\ 1998). We also measure the
concentration of these composite galaxies,
$C_{\eta}=F\,(<\theta_{0.5})\,/\,F\,(<1.5\,\theta_{0.5})$, i.e.\ the
ratio between the flux within the radius $\theta_{0.5}$ and that
within 1.5\,$\theta_{0.5}$ (Saracco et al.\ 1999).

%
%
\setcounter{figure}{5}
\medskip
\begin{figure*}
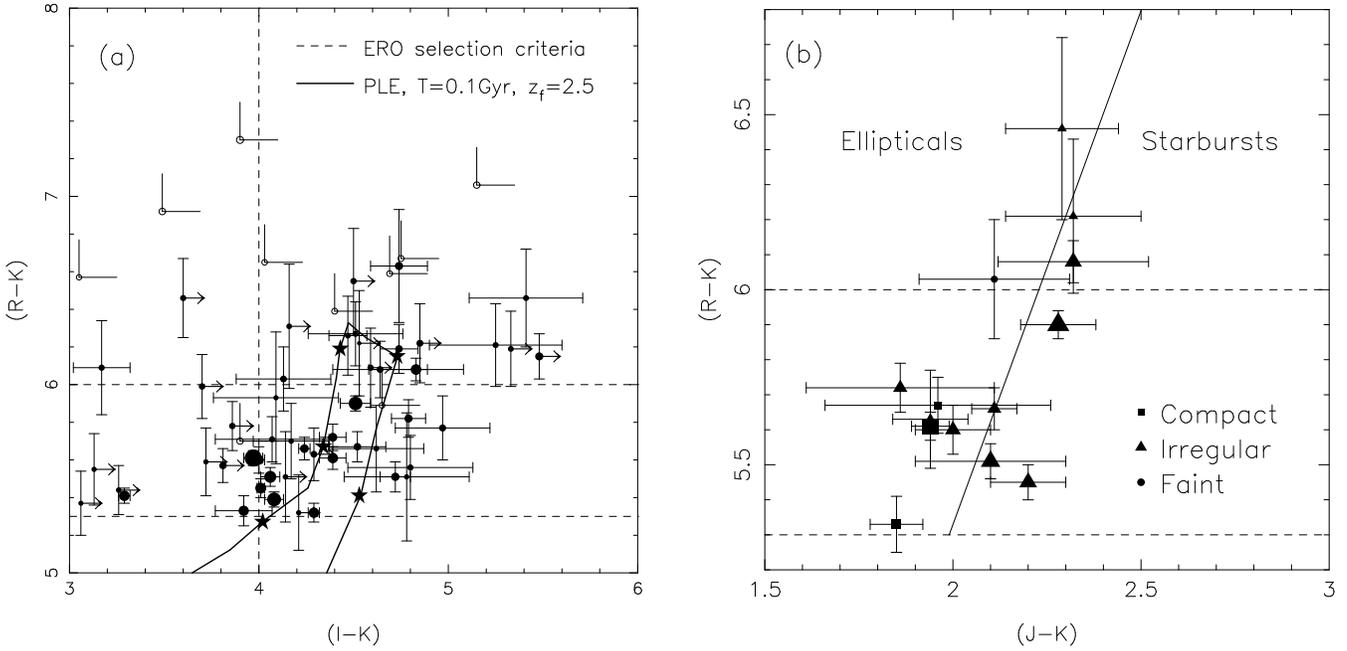

\centerline{\psfig{file=figs/fig6a.ps,angle=-90,width=85mm}
\hspace{5mm}
\psfig{file=figs/fig6b.ps,angle=0,width=85mm}}
\caption{a) $(R-K)$--$(I-K)$ colour-colour diagram showing our sample
of EROs with $(R-K)\ge5.3$. The predicted evolution of an early type
galaxy in the $z_f=2.5$, $\tau=0.1$\,Gyr PLE model (Daddi et al.\ 2000b)
is shown by the solid line. We mark redshift intervals of $\Delta
z\sim0.25$ by stars, starting with $z=1$ at the lower
left. The excess of galaxies with colours redder than the elliptical
galaxy evolutionary track suggests that the $(R-K)\ge6.0$ ERO
population may contain a significant fraction of dusty starburst
galaxies. The size of the filled points scale with source plane $K$--band
magnitude and open points represent galaxies for which only limits on
both colours are available. b) $(R-K)$--$(J-K)$ colour-colour 
diagram showing the 
EROs in our sample for which we have obtained $J$--band detections. The
solid diagonal line shows $(J-K)=0.34(R-K)+0.19$ from Pozzetti\,\&\,Mannucci
(2000). 
The dashed lines mark our ERO selection criteria.
The size of the points scale with source plane $K$--band
magnitude and the symbols denote the morphological classification of each
ERO from \S3.5. 
}
\end{figure*}
\smallskip

%
%
\setcounter{figure}{6}
\begin{figure*}
\centerline{\psfig{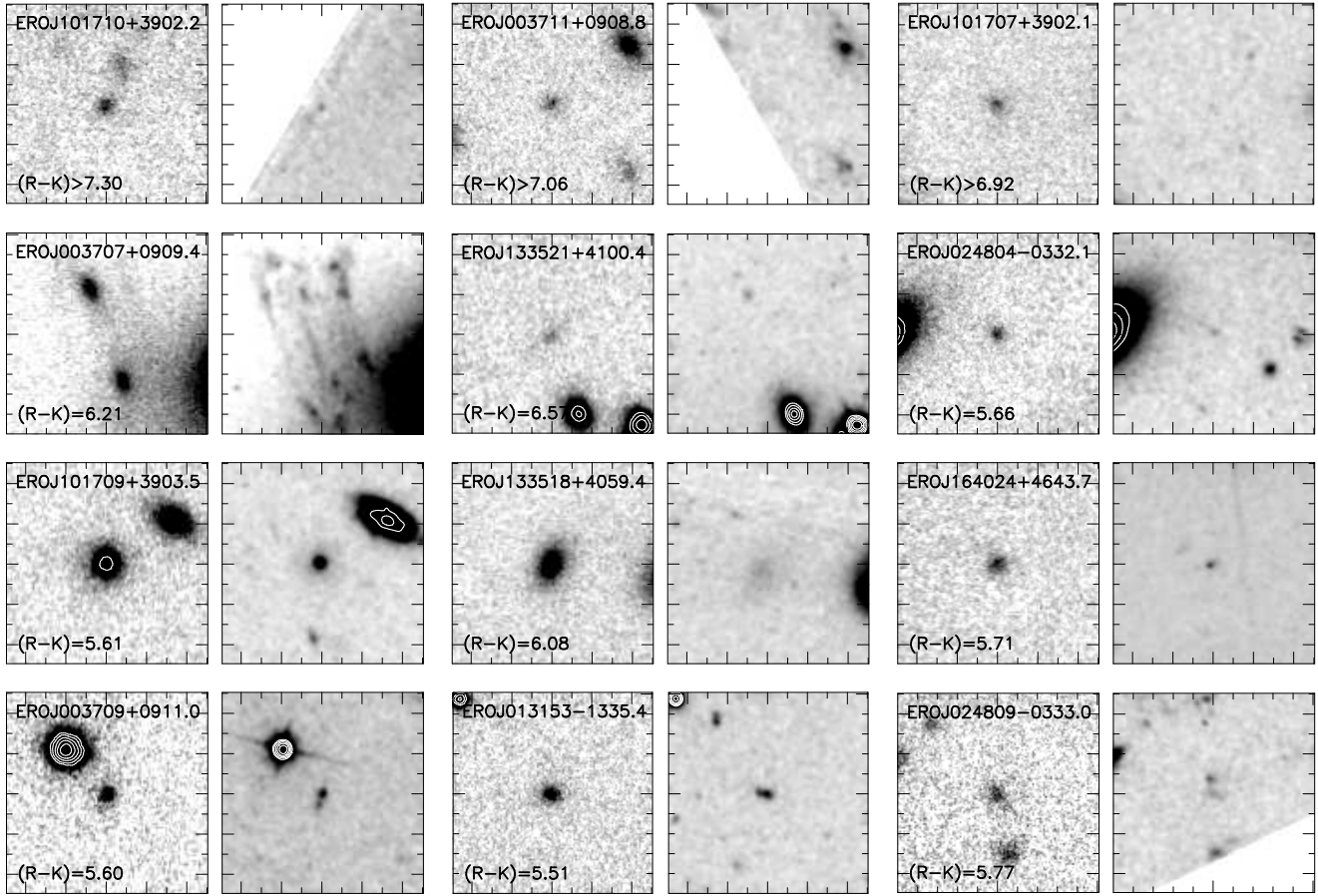}}
\noindent{\small\addtolength{\baselineskip}{-3pt}}
\caption{Images of a selection of EROs from the sample, showing the
$K$--band to the left and the F702W--band (smoothed with a 0.2$''$ FWHM
gaussian) to the right in each case. The top row shows the most
extreme EROs with $(R-K)\gs7$, the next row (upper middle) shows
examples of strongly lensed EROs, the third row (lower middle)
illustrates three of the more extended systems and the bottom row
gives apparently interacting or morphologically complex systems. Each
panel has North top and East left and is $10''$ square.}
\end{figure*}

Despite the co-addition of the data in these broad magnitude bins, no
discernible trend in scale-size or concentration with $K$--band
magnitude or colour is found; for example, the concentration values
differ by $\ls0.01$ while the error bars are typically $\gs0.05$.  It
therefore appears that, although our $K$--band data enjoys both high
signal to noise and high resolution (0.5$''$), a detailed
morphological analysis of these galaxies at near-infrared wavelengths
is not possible.  We therefore turn to the optical {\it HST} imaging,
noting that at $z\sim1$--2, the F702W filter samples the rest-frame
ultra-violet. The following morphological classification is therefore
sensitive to any ongoing unobscured star-formation in these galaxies, 
allowing us to identify easily starburst systems..

We classify our ERO sample on the basis of their appearance in the
{\it HST} F702W frames (Table~3): ``C'', compact galaxies; ``I'',
irregular galaxies, e.g.\ disk-like, clumpy or interacting morphology;
``F'', faint galaxies which are either not or only just detected in
the {\it HST} frames.  The $(R-K)\ge5.3$ sample comprises: 18\%\,(C);
50\%\,(I); 32\%\,(F).  Considering just the Compact and Irregular 
galaxies, we find that
$\sim90\%$ of the $(R-K)\ge6.0$ population are irregular, compared
with $\sim65\%$ of those with $5.3\le (R-K)\le6.0$.  This is another
hint that the $(R-K)\ge6.0$ population may be dominated by distant dusty
starbursts, although the large fraction of irregular $5.3\le
(R-K)\le6.0$ EROs, suggests that dusty starbursts may also make an
important contribution to the less extreme ERO population.

These results, specifically the large fraction of dusty starburst EROs
in our sample, disagree with previous attempts to classify EROs on the
basis of their morphology. For example, Moriondo et al.\ (2000) claim
that only $\sim$15--20\% of EROs in their sample have an ``irregular''
morphology, while 50--80\% are well fitted by an elliptical galaxy
($r^{1/4}$ law) profile. As noted by these authors, their data is
heterogeneous, being drawn from the {\it HST} archive, and so the
depth of their observations is not well defined, in contrast to our
highly homogeneous data set. Another difference between the two
studies is that $\sim65\%$ of Moriondo et al.'s EROs come from
targeted searches in probable dense environments, such as high
redshift clusters, and regions around radio galaxies and quasars, 
whereas we survey
random fields in the $z\gs1$ universe. Moriondo et al.'s (2000)
results may therefore be biased by their concentration on dense
environments.

Finally, we illustrate the diversity of ERO morphology in Fig.~7,
including candidate interacting galaxies and strongly lensed EROs.

\subsection{Clustering of EROs}

There is an order of magnitude variation in the number of EROs
detected in each cluster field (Fig.~2 \& Table~1), supporting
previous claims that at least some component of the ERO population is
strongly clustered (Daddi et al.\ 2000a).  However, to confidently
identify two or more EROs as lying at the same redshift and therefore
as being physically associated with each other generally requires
spectra of those galaxies. Here, we search our photometric catalogue
for examples of two or more neighbouring EROs that exhibit similar
$K$--band magnitudes and $(R-K)$ colours. The premise being that such
systems may represent high-redshift galaxy associations and that the
similar colours and magnitudes of such galaxies are analogous with the
strong early-type galaxy sequences observed in lower redshift clusters 
and groups (e.g.\ Fig.~2\,\&\,3a). We find two candidates for pairs of 
EROs, one each in A\,963 and A\,1835.

{\it EROJ101701$+$3903.4~~\&~~EROJ101703$+$3903.4} --- These two EROs
(Table~3) have $K\sim19$ and $(R-K)\sim6$ and lie north-west of the cD
galaxy in A\,963 in Fig.~1. They both have a regular $K$--band
morphology and very faint, low-surface brightness $R$-- and $J$--band
morphologies.  Within the photometric errors, the $(R-K)$ and $(J-K)$
colours of these two galaxies support the idea that these are two high
redshift elliptical galaxies. However,
ERO\,J101701$+$3903.4 is $\sim$0.5--1.0\,magnitude bluer than its
neighbour in $(I-K)$ (Fig.~5a \& Table~3), indicating that it may be
at a slightly lower redshift.

{\it EROJ140057$+$0252.4} --- This is one of two adjacent galaxies
(separation $\sim$3$''$) in the field of A\,1835. The other galaxy
falls just below the $(R-K)\ge5.3$ criterion (Fig.~2), and is
therefore not included in the ERO sample. Both galaxies have
featureless $K$--band morphologies, as does the ERO in the
$R$--band. The neighbour however has a very diffuse and low surface
brightness $R$--band morphology.

We conclude that there are no unambiguous examples of ERO 
associations in our sample.

\subsection{Strongly Lensed EROs}

The amplifying power of the cluster lenses becomes very high within
$\ls30''$ of the centre of the cluster. This is the region of the
image plane in which rare, highly magnified giant arcs are detected in
some clusters (e.g.\ A\,383, Smith et al.\ 2001a). In this same region
of the image plane, we detect three strongly lensed EROs in our
cluster sample, which we discuss below.

{\it EROJ003707$+$0909.4~~\&~~EROJ003707$+$0909.5} --- Three images of a
single background galaxy are detected in the core of A\,68. The two
brighter images are adjacent to the central galaxy and are shown in
Fig.~7, while the third image (ERO\,J0037006+0909.1) lies
$\sim20''$ south of the central galaxy and is
considerably fainter. The optical morphology of these images reveals
complex structure, containing what appears to be five bright knots
within each image. In contrast the $K$--band shows a bright centrally
concentrated source ($K$=17.2, 17.6 and 19.1 respectively). We obtain
colours of $(R-K)\sim6$, $(I-K)\sim5$ and $(J-K)\sim2.3$ for the central 
red region in both of the two brighter images (Table~3).  The morphology 
and position of these arcs relative to the central galaxy suggest that 
they are the images of a background spiral galaxy, probably at a redshift 
of $z\sim0.5$--1.0.

We also perform photometry using concentric circular apertures of
increasing diameter to estimate the colour gradient within the
brightest image. The colour gradient indicates that this galaxy only
has a colour of $(R-K)\ge5.3$ for photometric apertures of $\ls3''$ in
diameter. From our lens model we estimate that the amplification of 
the central region of the galaxy is $\sim15$, implying that a 
source plane aperture of
diameter $\ls0.75''$ would be required to identify this as an ERO.  
This source would therefore have $(R-K)<5.3$ (2$''$ diameter aperture 
-- see \S3.1) and would not be detected as as ERO in the absence of 
the foreground cluster lens.

{\it EROJ133521$+$4100.4} --- This ERO is $\sim20''$ north east of the
central galaxy in A\,1763 and is shown in Fig.~7. Its ellipticity
(a\,/\,b\,$\sim5$) and position angle with respect to the central
galaxy indicate that it is probably strongly lensed. We do not however
detect any counter images which implies that this galaxy is either
singly imaged, or that the counter images fall below the detection
threshold of our observations. The arclet has $K=19.8$ and
$(R-K)\sim6.6$ and we place lower limits on its $(I-K)$ and $(J-K)$
colours of $3.5$ and $1.5$ respectively, suggesting that this may be a
high redshift elliptical galaxy.

{\it EROJ024804$-$0332.1} --- This ERO was identified by Smith et al.\
(2001a) and lies $\sim25''$ south of the cD galaxy in A\,383 and
adjacent to a bright cluster elliptical.  The ellipticity 
(a\,/\,b\,$\sim$7) and position angle (tangential to the cD) of this 
ERO supporting the interpretation of this image as a lensed background 
galaxy.  We measure $(R-K)\sim6$
and $(I-K)\gs5$, making this an extreme ERO under both $(R-K)$ and 
$(I-K)$ selection criteria.  Smith et al.\ (2001a)
estimated from their lens model that the object's redshift is $z\ls4$.

\section{Discussion and Conclusions}

In this section we compare the observed number counts of EROs
with the theoretical predictions of pure luminosity evolution (Daddi et
al.\ 2000b; McCracken et al.\ 2000 -- hereafter 
M00) and semi-analytic (Cole et al.\ 2000 -- hereafter C00)
models of galaxy formation, and then summarise our conclusions. Our
primary aim when comparing our observational results with the different
model predictions is to investigate the suggestion that EROs comprise a
mixture of evolved and dusty galaxy populations.

\subsection{Comparison with PLE Models}

In Fig.~5 we compare our observed number counts (\S3.3) with Daddi et 
al.'s (2000b) PLE 
model down to $K\sim22$. This model attempts to describe EROs
as a single population of evolved galaxies using two free parameters:
$z_f$, the initial redshift of star formation in these galaxies and
$\tau$, the e-folding time of the starburst. Daddi et al.\ adopt a
Salpeter (1955) initial mass function (IMF), solar metallicity, no dust
reddening and normalise the predicted number counts to the local
luminosity function and rest frame colours of elliptical galaxies.
Looking at the range of model parameters in this figure we see that at
$K\sim21$ the models diverge, enabling us to discriminate between them.
For the $(R-K)\ge5.3$ sample, the models with $\tau=0.1$\,Gyr, $z_f=2.5$
and $\tau=0.3$\,Gyr, $z_f=3.5$, best match the data, with higher values
of $z_f$ predicting too many EROs at $K\gs20$. The same models are also
the best match with the $(R-K)\ge6.0$ sample.

However, as we show earlier, EROs appear not to be a single population 
of galaxies (\S1), with the fraction of dusty starbursts possibly increasing
towards redder optical/near-infrared colours and fainter $K$--band
magnitudes (\S3.3, \S3.5). Agreement between the observations and these
simple models should therefore deteriorate, at the faint, red limit. 
Unfortunately, comparison close to the faint limit is hampered by the 
cumulative nature of the data plotted in Fig.~5, fainter magnitudes 
being influenced by the brighter bins. 
Nevertheless, Fig.~5b suggests that the models have more
difficulty in correctly predicting the surface density of $(R-K)\ge6.0$
EROs, supporting the idea that this population contains a substantial
fraction of dusty starbursts. We also note that although the best model
lies within 1--$\sigma$ (Poissonian) of our data, the very small number
of these redder EROs in our sample, particularly at $K\sim18$--19,
implies that the disagreement between Daddi et al.'s (2000a) data and
this model at $K\sim18$--20 is a better test of the model at these
magnitudes.  Daddi et al.\ (2000b) also noticed this discrepancy between 
their $(R-K)\ge6$ data and the models, and suggested that it was due to a
possible deficit of $(R-K)\ge6.0$ ellipticals at $K\ls19.5$.

In summary, a simple model that assumes EROs comprise only passively
evolving elliptical galaxies (with $\tau=0.1$\,Gyr and $z_f=2.5$)
succeeds fairly well in predicting the number density of the
$(R-K)\ge5.3$ population. This model is also the best match to
observations of the $(R-K)\ge6.0$ population, however the agreement is
significantly worse, suggesting that the ``single population'' assumption
may be a poor description of these more extreme EROs.

We also compare our observations with a more general PLE model which
attempts to describe the whole galaxy population, and not just passive
ellipticals. This more detailed model (M00) includes
five galaxy populations (E/S0, Sab, Sbc, Scd, Sdm), each normalised to
observed local galaxy parameters (i.e. luminosity function and rest frame
colours). M00 also require their model to reproduce the
shape and amplitude of Cowie et al.'s (1996) $K$--selected redshift
distribution, which contains very few $K<19$ galaxies at $z>1$.  To 
achieve this, they adopt a dwarf dominated IMF ($x=3$) in order to 
reconcile their PLE prescription with Cowie et al.'s observational results.

The M00 model (Fig.~5) under-predicts the surface
density of $(R-K)\ge5.3$ EROs by approximately an order of magnitude and
predicts none with $(R-K)\ge6.0$.  We suggest that this is
probably caused by M00 requiring their model to fit Cowie et 
al.'s (1996) $K$--selected redshift distribution, which contains 
comparatively few galaxies at $z>1$.  However, Cowie et al.\ 
(1996) were concerned that their optical follow-up of a small-field 
$K$--selected sample might be incomplete for the reddest (i.e.\ optically 
faintest -- $R\gs24$) galaxies, and hence for galaxies at $z\gs1$.  
Recent wide field surveys have also discovered that EROs are strongly 
clustered (e.g.\ Daddi et al.\ 2000a: $\sim700$arcmin$^2$).  In contrast, 
Cowie et al.'s spectroscopic survey covered just 26.2\,arcmin$^2$, 
raising the possibility that their survey targeted an under-dense patch 
of sky.  We therefore suggest that near-infrared spectroscopy of 
wide field $K$--selected samples is necessary before the redshift 
distribution of such galaxies beyond $z\sim1$ can be reliably quantified.

\subsection{Comparison with Semi-analytic Models}

We now return to the primary aim of our comparison with model
predictions, that of investigating the suggestion that EROs comprise a
mixture of evolved and dusty galaxy populations.

We compare our ERO number counts with the predictions of the ``reference 
model'' from C00.  This is a semi-analytic model which 
calculates the formation and
evolution of galaxies in hierarchical clustering cosmologies, based on
N-body simulations and simple parametrisation of physical processes. The
model parameters are constrained by a number of local galaxy properties 
including the
ratio of elliptical to spiral galaxies, the metallicity of local $L^*$
ellipticals, $B$-- and $K$--band luminosity functions, the fraction of
gas in spiral and irregular galaxies and the size of galaxy disks. C00 
assume a Kennicutt (1993) IMF, however they find that the
optical/near-infrared colours of galaxies are insensitive to this choice.
They also include dust extinction using a Milky Way extinction curve, however
as the authors point out, they do not allow for clumping of the dust and
stars, nor do they include the effects of dust emission. We note that 
the model predicts a much weaker colour-magnitude
correlation for cluster elliptical galaxies than that observed in the
Coma cluster,  the predicted colour of cluster ellipticals being 
$\sim$0.1--0.4 magnitudes bluer than the observations.

The surface density of EROs predicted by the C00 reference
model is shown in Fig.~5 (Baugh, private communication). Despite its
success in reproducing the properties of local galaxies, this model
under-predicts the surface density of EROs with $(R-K)\ge5.3$ and 6.0 by
approximately an order of magnitude.  It would obviously be inappropriate 
to modify this model to fit the ERO number counts if in doing so, it no 
longer agreed with the local observational constraints mentioned above.  
Nevertheless, this disagreement with observations may point to important 
opportunities to improve our understanding of galaxy formation and hence 
improve C00's semi-analytic model.  We therefore briefly consider 
where these opportunities may lie.

As EROs appear to comprise galaxies containing both evolved and 
dust-reddened stellar populations (\S1, \S3.3 and \S3.5), C00's
under-prediction of EROs suggests that their reference model 
contains insufficient old stars and/or dust at $z\sim1$--2.  It therefore 
appears that their reference model does not produce enough stars and/or dust 
at high-redshift ($z\gg2$). One possible remedy would be to increase the 
fraction of stars formed in bursts, at the expense of quiescent star 
formation. This should reduce the level of ongoing star formation at
$z>1$ (which would otherwise make these galaxies too blue to be classed
as EROs). Another possible remedy would be to adopt a giant dominated IMF
as this would increase the quantity of dust produced through the
formation of a larger fraction of massive stars, thus both making
starburst galaxies redder and allowing more systems to go through a
starburst ERO phase without altering the properties of the local galaxy
population with which the current model agrees.  Introducing dust clumping
may also provide a more realistic description of the spatial distribution of
dust in starburst systems.  These options should be considered in future 
revisions of the semi-analytic models.

Although the C00 reference model fails to reproduce the observed
number of EROs, this model does provide a physically motivated framework
to qualitatively interpret the relationship between the various classes 
of ERO.  In the model the redshifts of galaxies predicted
to have $5.3\le (R-K)\le6.0$ and $K\ls22$ are $z\sim0.8$--2.5, equally
split between the $z\sim0.8$--1.5 and $z\sim1.5$--2.5 bins. The lower
redshift bin is dominated by passively evolving galaxies ($\sim90\%$ of
the total) whose stars are sufficiently old to produce such red colours
in the observed passbands, whilst the higher redshift bin is dominated by
starburst galaxies ($\sim95\%$ of the total), whose $(R-K)$ colours are
reddened by dust. A similar analysis of the predicted $(R-K)\ge6.0$
population reveals that almost all of these more extreme EROs are
predicted to lie at $z\sim1.5$--2.5, with $\sim70\%$ of these galaxies
being dusty starbursts. The C00 reference model therefore
appears to qualitatively support the observational evidence (\S3) for
both a predominance of distant dusty starburst galaxies in the
$(R-K)\ge6.0$ EROs, and passively evolving galaxies at $z\sim1$
producing the bulk of the $(R-K)\sim5.3$--6.0 population.

We now return to the break in the number counts of EROs identified in
\S3. Whilst the normalisation of the semi-analytic predictions falls an
order of magnitude short of our observations, the qualitative properties
of the predicted ERO population offers a plausible explanation of the
break in the count slopes. At magnitudes brighter than the break, the
counts may be dominated by lower redshift ($z\sim0.8$--1.5),
predominantly passively evolving galaxies, whereas faint-ward of the
break, the counts could be dominated by the distant ($z\sim1.5$--2.5),
dusty starburst galaxies. We therefore speculate that the break may be
due to a transition from an ERO population dominated by evolved galaxies
at $z\sim1$ ($K\ls19.5$) to one dominated by distant dusty starburst
galaxies (that may have experienced a recent merger) at $z\sim2$
($K\gs19.5$).

\subsection{Conclusions}

We have undertaken a deep optical/near-infrared survey of 10 massive
cluster lenses at $z\sim0.2$ using {\it HST} and UKIRT. We find 60
EROs with $(R-K)\geq 5.3$, of which 26 have $(R-K)\geq 6.0$ in a total
image plane survey area of 49 arcmin$^2$ down to $K=20.6$.

We use detailed models of the cluster lenses (S01) to quantify the 
lens amplification and thus correct the observed number counts for the 
effects of gravitational lensing.  After making these corrections, we 
estimate a surface density of $2.5\pm0.4$\,($1.2\pm0.3$)
arcmin$^{-2}$ for EROs with $(R-K)\geq 5.3$\,($6.0$) at
$K\le21.6$. Our results agree with previous shallower wide-field
surveys at $K\ls19$ and probe the number density of EROs with
$(R-K)\ge5.3$ and 6.0 down to a source plane magnitude of $K\sim22$.

The number counts of both classes of ERO flatten significantly at
magnitudes fainter than $K\sim19$--20.  We speculate that this is 
due to a transition from an ERO population dominated by evolved 
galaxies at
$z\sim1$--2 ($K\ls19$--20) to one dominated by distant dusty starburst
galaxies at $z>1$ ($K\gs19$--20). Analysis of the $(R-K)$--$(I-K)$
and $(R-K)$--$(J-K)$ planes also suggests that the $(R-K)\ge6.0$
population may contain a substantial fraction of dusty starburst
galaxies.

Approximately 50\% of our sample contain morphological substructure
including disk-like, clumpy or interacting morphologies. This is a
larger fraction than found by previous studies (e.g.\ Moriondo et al.\
2000) which claimed that only $\sim$15--20\% of EROs have such
morphologies. The discrepancy may be due to a bias towards dense
environments and the heterogeneity of Moriondo et al.'s dataset,
compared with our unbiased and more homogeneous dataset.

We compare our observations with progressively more sophisticated
models of galaxy formation, beginning with a two parameter ($z_f$ and
$\tau$) PLE model that attempts to describe EROs as a single
population of elliptical galaxies (Daddi et al.\ 2000b). The model
parameters which best match the observations are $\tau=0.1$\,Gyr and
$z_f=2.5$, ruling out the very high formation redshifts ($z_f\sim10$)
that were allowed by Daddi et al.'s (2000a) shallower
observations. However, this single population model matches the
$(R-K)\ge6.0$ EROs significantly worse than the $(R-K)\ge5.3$ EROs,
supporting the idea that the more extreme population contains a large
fraction of distant dusty starbursts in addition to the elliptical
galaxies contained within this model.

We then compared our observations with PLE models that attempt to
describe the whole galaxy population, and not just passive ellipticals
(M00).  These models under-predict the surface density of $(R-K)\ge5.3$
EROs by approximately an order of magnitude and predict none with
$(R-K)\ge6.0$. This deficit of EROs is probably caused by M00 requiring their 
model to fit the median redshift of Cowie et al.'s (1996) $K$--selected 
redshift distribution.  We suggest that this confirms Cowie 
et al.'s (1996) concern that their optical follow-up of a small-field 
$K$--selected sample might be incomplete for the reddest (i.e.\ optically 
faintest -- $R\gs24$) galaxies, and hence for galaxies at $z\gs1$.  
It therefore appears that near-infrared spectroscopy of wide field 
$K$--selected samples is necessary before the redshift distribution 
of galaxies at $z\gs1$ can be reliably quantified.

Finally, we compare the observed number density of EROs with 
the semi-analytic predictions from the reference model of C00. 
This semi-analytic model under-predicts the number density of EROs 
at $K\sim18$--22 by an order of magnitude, indicating that the 
current generation of semi-analytic models may not produce sufficient 
stars and/or dust at high redshift ($z>2$).  However, as the 
C00 reference model is physically well motivated, we look at the 
predicted properties of the ERO population, finding that their 
redshift distribution and the split between passive and dusty 
EROs appear to support our interpretation of the break in the 
slope of the number counts.

This is the first survey to exploit massive foreground galaxy clusters
to amplify the flux of background EROs. The sample constructed from
our deep, high resolution $K$-- and $R$--band observations is
therefore ideally suited to near-infrared spectroscopic follow-up on
10--m class telescopes (e.g.\ Smith et al.\ 2001b), the lens 
amplification allowing us to probe
$\sim1$ magnitude beyond the normal reach of such observations.
The primary goal of our future programme will be to probe the properties of 
EROs beyond the break in their number counts at $K\sim19$--20 and to 
identify the cause of this feature.

\section*{Acknowledgements}

We thank the referee, Alfonso Arag\'on-Salamanca for his careful reading 
of the manuscript and numerous thoughtful comments.  
We also thank Richard Ellis, Andy Fruchter, Richard Hook, Harald
Kuntschner, Peder Norberg and Kevin Pimbblet for useful discussions
and assistance at various stages of this project.
We are also very grateful to Carlton Baugh, Shaun Cole, Emanuele Daddi, 
Nigel Metcalfe and Tom Shanks for providing the theoretical predictions
discussed in this paper.
Finally, we thank Andy Adamson, John Davies and Tom Kerr for support of
the UKIRT observing programme and recovery observations.

GPS acknowledges a postgraduate studentship from PPARC.
IRS acknowledges support from the Royal Society and the
Leverhulme Trust.
JPK acknowledges support from CNRS.
OC acknowledges support from the European Commission under
contract no.\ ER-BFM-BI-CT97-2471.
HE acknowledges support by NASA and STScIgrants NAG 5--6336 and
GO 5--08249.
RJI acknowledges support from PPARC.
We also acknowledge financial support from the UK--French
ALLIANCE collaboration programme \#00161XM

\newpage

%
%
{\small
\setcounter{table}{2}
\begin{table*}
\begin{center}
\caption{\hfil Catalogue of EROs \hfil}
\begin{tabular}{lccccccl}
\noalign{\smallskip}\hline
\noalign{\medskip}
{Source} & {$\alpha$,$\delta$} & $K$ & $(R-K)$ & $(I-K)$ &
$(J-K)$ & FWHM &
Morphology \cr
& (J2000) & & & & & ($''$) & (a)\cr
\noalign{\smallskip}\hline
\noalign{\smallskip}
\multispan2{A\,68 \hfil} \cr
\noalign{\smallskip}
ERO\,J003703+0909.7 & 00\,37\,03.30 $+$09\,09\,44.4 &
$20.10\pm0.16$ & $>6.39$ & $>4.40$ & $>1.42$ & $1.23$ & F \cr 
ERO\,J003706+0909.1 & 00\,37\,06.10 $+$09\,09\,08.7 &
$19.10\pm0.01$ & $>5.70$ & $>3.86$ & $>2.06$ & $0.92$ & I, \S3.6\cr
ERO\,J003707+0909.4 & 00\,37\,07.25 $+$09\,09\,23.8 &
$17.62\pm0.03$ & $6.21\pm0.22$ & $5.25\pm0.35$ & $2.32\pm0.18$ & $0.62$ &
I, \S3.6\cr 
ERO\,J003707+0909.5 & 00\,37\,07.37 $+$09\,09\,28.4 &
$17.16\pm0.03$ & $6.46\pm0.26$ & $5.41\pm0.30$ & $2.29\pm0.15$ & $0.66$ &
I, \S3.6\cr 
ERO\,J003709+0909.6 & 00\,37\,08.84 $+$09\,09\,37.0 &
$19.45\pm0.10$ & $6.19\pm0.20$ & $>5.33$ & $>2.07$ & $0.25$ & F \cr
ERO\,J003709+0911.0 & 00\,37\,09.03 $+$09\,11\,01.8 &
$18.21\pm0.04$ & $5.60\pm0.07$ & $4.00\pm0.03$ & $2.00\pm0.10$ & $3.86$ &
I, Clumpy \cr 
ERO\,J003710+0908.9 & 00\,37\,09.57 $+$09\,08\,53.4 &
$18.17\pm0.03$ & $5.66\pm0.06$ & $4.24\pm0.03$ & $2.11\pm0.06$ & $0.18$ &
I \cr 
ERO\,J003710+0909.1 & 00\,37\,09.58 $+$09\,09\,08.4 &
$17.53\pm0.03$ & $5.45\pm0.05$ & $4.01\pm0.02$ & $2.20\pm0.10$ & $0.45$ &
I, Disk?\cr 
ERO\,J003711+0908.8 & 00\,37\,10.53 $+$09\,08\,46.8 &
$19.81\pm0.12$ & $>7.06$ & $>5.15$ & $>1.71$ & $0.46$ & F \cr
ERO\,J003711+0909.1 & 00\,37\,10.91 $+$09\,09\,09.8 &
$20.17\pm0.16$ & $>6.67$ & $>4.75$ & $>1.35$ & $0.27$ & F \cr 
ERO\,J003711+0909.2 & 00\,37\,11.00 $+$09\,09\,10.1 &
$20.30\pm0.17$ & $>6.59$ & $>4.69$ & $>1.22$ & $0.89$ & F \cr 
\noalign{\medskip}
\multispan2{A\,209 \hfil} \cr
\noalign{\smallskip}
ERO\,J013151$-$1335.5 & 01\,31\,50.78 $-$13\,35\,31.0 &
$20.07\pm0.15$ & $5.70\pm0.20$ & $4.17\pm0.20$ & ... & $0.34$ &
C \cr 
ERO\,J013153$-$1335.4 & 01\,31\,52.82 $-$13\,35\,22.8 &
$18.65\pm0.05$ & $5.51\pm0.08$ & $4.72\pm0.08$ & ... & $0.39$ &
I, Clumpy \cr 
ERO\,J013154$-$1336.3 & 01\,31\,54.14 $-$13\,36\,19.9 &
$19.54\pm0.11$ & $6.26\pm0.21$ & $4.47\pm0.10$ & ... & $0.42$ &
F \cr 
ERO\,J013157$-$1336.6 & 01\,31\,56.98 $-$13\,36\,37.5 &
$17.98\pm0.03$ & $5.32\pm0.05$ & $4.29\pm0.03$ & ... & $0.50$ &
I \cr 
ERO\,J013159$-$1336.2 & 01\,31\,58.60 $-$13\,36\,15.4 &
$20.06\pm0.15$ & $5.66\pm0.23$ & $4.62\pm0.25$ & ... & $0.31$ &
C \cr 
ERO\,J013159$-$1336.3 & 01\,31\,59.36 $-$13\,36\,16.4 &
$18.72\pm0.06$ & $5.82\pm0.10$ & $4.79\pm0.09$ & ... & $0.33$ &
I, Disk\cr 
ERO\,J013200$-$1336.5 & 01\,31\,59.66 $-$13\,36\,28.4 &
$18.96\pm0.09$ & $6.19\pm0.13$ & $4.74\pm0.10$ & ... & $0.36$ &
F \cr 
\noalign{\medskip}
\multispan2{A\,267 \hfil} \cr
\noalign{\smallskip}
ERO\,J015238+0101.0 & 01\,52\,38.43 $+$01\,02\,02.8 &
$17.76\pm0.02$ & $5.51\pm0.05$ & $4.06\pm0.05$ & $2.10\pm0.20$ & $0.42$ &
I, Disk? \cr 
ERO\,J015238+0101.9 & 01\,52\,37.60 $+$01\,01\,51.6 &
$18.73\pm0.05$ & $6.15\pm0.12$ & $>5.48$ & $>1.85$ & $0.67$ & C \cr
ERO\,J015240+0101.2 & 01\,52\,40.08 $+$01\,01\,08.8 &
$18.74\pm0.05$ & $5.67\pm0.08$ & $4.52\pm0.15$ & $1.96\pm0.30$ & $0.35$ &
C \cr 
ERO\,J015240+0101.6 & 01\,52\,39.55 $+$01\,00\,38.4 &
$20.37\pm0.17$ & $>6.65$ & $>4.03$ & $>0.20$ & $0.57$ & F \cr 
ERO\,J015240+0101.7 & 01\,52\,39.88 $+$01\,01\,39.3 &
$20.32\pm0.15$ & $5.93\pm0.35$ & $4.09\pm0.33$ & $>0.26$ & $0.34$ &
C \cr 
ERO\,J015240+0101.8 & 01\,52\,39.82 $+$01\,01\,50.0 &
$19.51\pm0.10$ & $6.09\pm0.21$ & $>4.59$ & $>1.07$ & $0.53$ & F \cr
ERO\,J015241+0101.1 & 01\,52\,41.25 $+$01\,02\,07.4 &
$19.14\pm0.09$ & $6.22\pm0.21$ & $>4.85$ & $>1.44$ & $0.68$ & F \cr
ERO\,J015243+0101.5 & 01\,52\,42.80 $+$01\,01\,27.0 &
$19.93\pm0.11$ & $5.51\pm0.34$ & $4.78\pm0.33$ & $>0.65$ & $1.30$ &
C \cr 
ERO\,J015245+0101.8 & 01\,52\,44.53 $+$01\,00\,47.9 &
$20.04\pm0.16$ & $5.32\pm0.20$ & $>4.21$ & $>0.54$ & $0.60$ & F
\cr
ERO\,J015247+0101.1 & 01\,52\,47.28 $+$01\,01\,04.8 &
$18.48\pm0.03$ & $5.72\pm0.07$ & $4.39\pm0.07$ & $1.86\pm0.25$ & $0.29$ &
I \cr 
ERO\,J015249+0100.6 & 01\,52\,48.93 $+$01\,00\,38.1 &
$20.58\pm0.16$ & $>5.89$ & $>4.65$ & $>0.03$ & $0.51$ & F \cr 
ERO\,J015249+0101.6 & 01\,52\,49.10 $+$01\,00\,36.8 &
$20.05\pm0.11$ & $5.51\pm0.24$ & $>4.14$ & $>0.53$ & $0.52$ & F \cr
\noalign{\medskip}
\multispan2{A\,383 \hfil} \cr
\noalign{\smallskip}
ERO\,J024804$-$0332.1 & 02\,48\,04.12 $-$03\,32\,06.4 &
$19.73\pm0.07$ & $6.22\pm0.28$ & $>4.53$ & ... & $0.29$ & C,
\S3.6\cr 
ERO\,J024805$-$0330.0 & 02\,48\,04.55 $-$03\,29\,57.3 &
$19.59\pm0.09$ & $6.27\pm0.17$ & $4.51 \pm0.25$ & ... & $0.46$ &
I \cr 
ERO\,J024805$-$0330.2 & 02\,48\,04.56 $-$03\,30\,11.0 &
$18.12\pm0.02$ & $5.41\pm0.04$ & $3.29\pm0.03$ & ... & $0.00$ &
C, Star? \cr 
ERO\,J024805$-$0330.3 & 02\,48\,04.82 $-$03\,30\,18.0 &
$18.06\pm0.04$ & $5.61\pm0.06$ & $4.39\pm0.07$ & ... & $0.47$ &
I \cr 
ERO\,J024805$-$0330.3 & 02\,48\,04.90 $-$03\,30\,22.7 &
$20.06\pm0.11$ & $6.31\pm0.33$ & $>4.16$ & ... & $0.36$ &
C \cr 
ERO\,J024805$-$0330.4 & 02\,48\,04.89 $-$03\,30\,24.7 &
$19.18\pm0.08$ & $6.08\pm0.15$ & $4.64\pm0.25$ & ... & $0.33$ &
I \cr 
ERO\,J024806$-$0331.5 & 02\,48\,06.44 $-$03\,31\,27.3 &
$18.27\pm0.05$ & $6.63\pm0.30$ & $4.74 \pm0.15$ & ... & $0.54$ &
I \cr 
ERO\,J024808$-$0331.8 & 02\,48\,07.99 $-$03\,31\,47.5 &
$19.44\pm0.10$ & $6.55\pm0.28$ & $>4.50$ & ... & $0.63$ &
I \cr 
ERO\,J024809$-$0332.9 & 02\,48\,08.79 $-$03\,32\,56.4 &
$19.47\pm0.11$ & $5.56\pm0.17$ & $4.80\pm0.33$ & ... & $0.62$ &
F \cr 
ERO\,J024809$-$0333.0 & 02\,48\,08.58 $-$03\,33\,00.0 &
$19.00\pm0.05$ & $5.77\pm0.17$ & $4.97\pm0.25$ & ... & $0.61$ &
I, Merging?\cr 
ERO\,J024810$-$0332.4 & 02\,48\,10.02 $-$03\,32\,25.8 &
$17.57\pm0.02$ & $5.39\pm0.04$ & $4.08\pm0.05$ & ... & $0.61$ &
I \cr 
\noalign{\medskip}
\multispan2{A\,773 \hfil} \cr
\noalign{\smallskip}
ERO\,J091802+5143.1 & 09\,18\,02.50 $+$51\,43\,06.4 &
$20.23\pm0.11$ & $5.59\pm0.18$ & $>3.73$ & $>1.23$ & $0.34$ & I,
Disk \cr 
\noalign{\medskip}
\multispan2{A\,963 \hfil} \cr
\noalign{\smallskip}
ERO\,J101701+3903.4 & 10\,17\,01.42 $+$39\,03\,24.6 &
$19.32\pm0.05$ & $6.09\pm0.25$ & $3.17\pm0.15$ & $>2.13$ &
$0.47$ & I \cr 
ERO\,J101703+3903.4 & 10\,17\,02.52 $+$39\,03\,22.0 &
$18.84\pm0.04$ & $6.03\pm0.17$ & $4.13\pm0.25$ & $2.11\pm0.11$ &
$0.58$ & F \cr 
ERO\,J101706+3901.7 & 10\,17\,05.69 $+$39\,01\,41.6 &
$20.53\pm0.15$ & $6.38\pm0.34$ & $>2.94$ & $>0.80$ & $0.00$ & F
\cr 
ERO\,J101706+3902.0 & 10\,17\,05.79 $+$39\,02\,02.3 &
$20.16\pm0.10$ & $5.55\pm0.19$ & $>3.13$ & $>1.04$ &
$0.57$ & F \cr 
ERO\,J101707+3901.4 & 10\,17\,06.61 $+$39\,01\,22.1 &
$19.51\pm0.08$ & $5.78\pm0.13$ & $>3.86$ & $>1.71$ & $0.11$ & I
\cr 
ERO\,J101707+3902.1 & 10\,17\,06.58 $+$39\,02\,08.3 &
$19.74\pm0.09$ & $>6.90$ & $>3.49$ & $>1.27$ & $0.21$ & I,
Clumpy \cr 
ERO\,J101707+3902.4 & 10\,17\,07.09 $+$39\,02\,25.0 &
$20.37\pm0.09$ & $5.37\pm0.17$ & $>3.06$ & $>0.96$ &
$0.28$ & I \cr 
\noalign{\smallskip}\hline
\noalign{\smallskip}
\end{tabular}
\end{center}
{\small (a) ``C'', Compact; ``I'', Irregular; ``F'', Faint morphological 
classification, as discussed in \S3.}
\end{table*}
}
{\small
\setcounter{table}{2}
\begin{table*}
\begin{center}
\caption{\hfil (continued) Catalogue of EROs \hfil}
\begin{tabular}{lccccccl}
\noalign{\smallskip}\hline
\noalign{\medskip}
{Source} & {$\alpha$,$\delta$} & $K$ & $(R-K)$ & $(I-K)$ &
$(J-K)$ & FWHM
& Morphology \cr
& (J2000) & & & & & ($''$) & (a)\cr
\noalign{\smallskip}\hline
\noalign{\smallskip}
\multispan2{A\,963 (continued) \hfil} \cr
\noalign{\smallskip}
ERO\,J101707+3902.9 & 10\,17\,07.43 $+$39\,02\,52.2 &
$19.14\pm0.04$ & $5.63\pm0.14$ & $>4.29$ & $1.94\pm0.10$ &
$0.46$ & I, Disk? \cr 
ERO\,J101709+3903.5 & 10\,17\,09.30 $+$39\,03\,31.4 &
$17.15\pm0.01$ & $5.61\pm0.04$ & $3.97\pm0.05$ & $1.94\pm0.05$ &
$0.56$ & C \cr 
ERO\,J101710+3902.2 & 10\,17\,09.66 $+$39\,02\,10.0 &
$19.60\pm0.08$ & $>7.28$ & $>3.90$ & $>1.65$ & $0.64$& I
\cr 
ERO\,J101710+3903.0 & 10\,17\,09.61 $+$39\,03\,03.6 &
$19.49\pm0.07$ & $5.99\pm0.17$ & $>3.70$ & $>1.55$ &
$1.20$ & I \cr 
ERO\,J101710+3903.4 & 10\,17\,09.94 $+$39\,03\,23.3 &
$19.66\pm0.09$ & $6.46\pm0.21$ & $>3.60$ & $>1.47$ & $0.85$ & F
\cr 
\noalign{\medskip}
\multispan2{A\,1763 \hfil} \cr
\noalign{\smallskip}
ERO\,J133511+4100.3 & 13\,35\,11.46 $+$41\,00\,15.5 &
$18.92\pm0.06$ & $5.57\pm0.09$ & $>3.81$ & ... & $1.25$ & I \cr
ERO\,J133517+4058.7 & 13\,35\,16.77 $+$40\,58\,43.0 &
$20.11\pm0.15$ & $5.44\pm0.13$ & $>3.26$ & $>1.57$ & $0.63$ & I
\cr 
ERO\,J133518+4058.8 & 13\,35\,17.82 $+$40\,58\,45.9 &
$20.47\pm0.16$ & $5.42\pm0.20$ & $>2.70$ & $>0.98$ & $0.64$ & I
\cr 
ERO\,J133518+4059.4 & 13\,35\,18.30 $+$40\,59\,25.8 &
$17.65\pm0.02$ & $6.08\pm0.06$ & $4.83\pm0.25$ & $2.32\pm0.20$ &
$0.76$ & I \cr 
ERO\,J133521+4100.4 & 13\,35\,21.16 $+$41\,00\,24.4 &
$19.84\pm0.10$ & $6.57\pm0.51$ & $>3.05$ & $>1.05$ & $1.24$ & F,
\S3.6\cr 
\noalign{\medskip}
\multispan2{A\,1835 \hfil} \cr
\noalign{\smallskip}
ERO\,J140057+0252.4 & 14\,00\,57.13 $+$02\,52\,26.4 &
$18.01\pm0.02$ & $5.33\pm0.08$ & $3.92\pm0.15$ & $1.85\pm0.07$ &
$0.53$ & C \cr 
\noalign{\medskip}
\multispan2{A\,2219 \hfil} \cr
\noalign{\smallskip}
ERO\,J164023+4644.0 & 16\,40\,23.05 $+$46\,44\,02.3 &
$17.51\pm0.01$ & $5.90\pm0.04$ & $ 4.51\pm0.08$ & $2.28\pm0.10$ & $0.51$ &
I, Disk \cr 
ERO\,J164024+4643.7 & 16\,40\,23.95 $+$46\,43\,42.1 &
$19.64\pm0.05$ & $5.71\pm0.12$ & $4.07\pm0.30$ & $>1.76$ & $0.68$ &
I \cr 
\noalign{\smallskip}\hline
\noalign{\smallskip}
\end{tabular}
\end{center}
{\small (a) ``C'', Compact; ``I'', Irregular; ``F'', Faint morphological 
classification, as discussed in \S3.}
\end{table*}
}

\pagebreak
\end{document}